\documentclass[journal, twoside, final]{IEEEtran}
\usepackage{cite}
\usepackage{graphicx}
\usepackage{amssymb, amsmath, amsthm}
\usepackage{amsfonts}
\usepackage{multirow}
\usepackage{mathrsfs}
\usepackage{color}
\usepackage{url}
\usepackage{flushend}
\usepackage{setspace}
\usepackage{placeins}
\usepackage{booktabs, multicol, multirow}
\usepackage{acronym}
\usepackage{bm}
\usepackage{placeins}
\usepackage{extarrows}
\usepackage{color}
\allowdisplaybreaks

\usepackage[caption=false]{subfig}

\usepackage{algorithm, algorithmic}

\usepackage{booktabs, multicol, multirow}
\usepackage{acronym}
\usepackage{bm}

\newtheorem{lemma}{Lemma}
\newtheorem{theorem}{Theorem}
\newtheorem{remark}{Remark}

\newcommand{\RNum}[1]{\uppercase\expandafter{\romannumeral #1\relax}}

\newmuskip\pFqmuskip
\newcommand*\pFq[6][8]{%
  \begingroup 
  \pFqmuskip=#1mu\relax
  \mathcode`\,=\string"8000
  \begingroup\lccode`\~=`\,
  \lowercase{\endgroup\let~}\pFqcomma
  {}_{#2}F_{#3}{\left[\genfrac..{0pt}{}{#4}{#5};#6\right]}%
  \endgroup
}
\newcommand{\pFqcomma}{\mskip\pFqmuskip}

\begin{document}

\title{Air-to-Air Communications Beyond 5G:\\A Novel 3D CoMP Transmission Scheme}

\author{Yan Li, Nikolaos I. Miridakis,~\IEEEmembership{Senior Member,~IEEE}, Theodoros A. Tsiftsis,~\IEEEmembership{Senior Member,~IEEE}, \\ Guanghua Yang,~\IEEEmembership{Senior Member,~IEEE}, and Minghua Xia,~\IEEEmembership{Member,~IEEE}
\thanks{Manuscript received February 12, 2020; revised May 26, 2020 and July 12, 2020; accepted July 12, 2020. This work was supported in part by the National Natural Science Foundation of China under Grant 61671488, in part by the China Postdoctoral Science Foundation under Grant 2019M653177, in part by the Major Science and Technology Special Project of Guangdong Province under Grant 2018B010114001, in part by the Fundamental Research Funds for the Central Universities under Grant 191gjc04, in part by the Science and Technology Planning Project of Guangdong Province under Grant 2019B010137006, and in part by the Guangzhou Leading Innovation Team Program (China) under Grant 201909010006. The associate editor coordinating the review of this paper and approving it for publication was W. Zhang.}
	\thanks{%
		Yan Li and Minghua Xia are with the School of Electronics and Information Technology, Sun Yat-sen University, Guangzhou 510006, China. Minghua Xia is also with the Southern Marine Science and Engineering Guangdong Laboratory, Zhuhai 519082, China (e-mail: liyan228@mail2.sysu.edu.cn; xiamingh@mail.sysu.edu.cn).
		
		Nikolaos I. Miridakis, Theodoros A. Tsiftsis and Guanghua Yang are with the school of Intelligent Systems Science and Engineering, and with the Institute of Physical Internet, Jinan University, Zhuhai 519070, China. Nikolaos I. Miridakis is also with the Department of Informatics and Computer Engineering, University of West Attica, Aegaleo 12243, Greece (e-mail: nikozm@uniwa.gr; \{theo\_tsiftsis, ghyang\}@jnu.edu.cn).}

	\thanks{%
		Color versions of one or more of the figures in this article are available online at https://ieeexplore.ieee.org.

		Digital Object Identifier XXX}
}

\markboth{IEEE Transactions on Wireless Communications} {Li \MakeLowercase{\textit{et al.}}: Air-to-Air Communications Beyond 5G: A Novel 3D CoMP Transmission Scheme}

\maketitle

\IEEEpubid{\begin{minipage}{\textwidth} \ \\[12pt] \centering 1536-1276 \copyright\ 2020 IEEE. Personal use is permitted, but republication/redistribution requires IEEE permission. \\
See \url{https://www.ieee.org/publications/rights/index.html} for more information.\end{minipage}}

\IEEEpubidadjcol

\begin{abstract}
\noindent In this paper, a novel $3$D cellular model consisting of aerial base stations (aBSs) and aerial user equipments (aUEs) is proposed, by integrating the coordinated multi-point (CoMP) transmission technique with the theory of stochastic geometry. For this new $3$D architecture, a tractable model for aBSs' deployment based on the binomial-Delaunay tetrahedralization is developed, which ensures seamless coverage for a given space. In addition, a versatile and practical frequency allocation scheme is designed to eliminate the inter-cell interference effectively. Based on this model, performance metrics including the achievable data rate and coverage probability are derived for two types of aUEs: {\it i)} the general aUE (i.e., an aUE having distinct distances from its serving aBSs) and {\it ii)} the worst-case aUE (i.e., an aUE having equal distances from its serving aBSs). Simulation and numerical results demonstrate that the proposed approach emphatically outperforms the conventional binomial-Voronoi tessellation without CoMP. Insightfully, it provides a similar performance to the binomial-Voronoi tessellation which utilizes the conventional CoMP scheme; yet, introducing a considerably reduced computational complexity and backhaul/signaling overhead. 
\end{abstract}


\section{Introduction}
\label{Section:Introduction}
\IEEEPARstart{W}{ith} the ever-increasing number of unmanned aerial vehicles (UAVs), drone-based wireless communications are expected to play a pivotal role to the establishment of upcoming future networking infrastructures \cite{8660516}. Due to their inherent mobility and flexibility, they can act as aerial base stations (aBSs) and aerial user equipments (aUEs). The former nodes can be used to support wireless connectivity for existing terrestrial base stations (BSs) by providing reliable and cost-effective on-the-fly communication. Also, they can offer additional throughput and coverage in some hotspots or to assist in emergency, disaster and critical situations. The latter nodes essentially act as mobile terminals or relays under the Internet of Things (IoT) and the emerging Internet-of-Space applications, by improving the connectivity and coverage of ground devices. They can also play a key role in surveillance and package delivery. Despite such attractive opportunities for UAVs, wireless communications via aerial nodes still face a number of challenges. Regarding aBSs, a key challenge is the three-dimensional (3D) deployment as well as other problems including the resource allocation, trajectory optimization, air-to-air/ground channel modeling, and performance analysis of aBS-enabled networks. Regarding aUEs, the key challenge is how to get reliable and high-throughput communication since the existing terrestrial BSs are mainly designed for serving terrestrial UEs, whereas they fail in meeting the high capacity/coverage demands of aUEs. Therefore, using aBSs to serve aUEs is quite promising and presents a requisite for future $5$G-and-beyond networking setups \cite{8869705}.

\IEEEpubidadjcol
\subsection{Related Works and Motivation}
There are a number of works focusing on wireless communications via UAVs. The theoretical investigation considering aUEs can be traced back to \cite{8269068}, where the coverage probability was analyzed. Another performance metric, the downlink achievable data rate for aUEs served by terrestrial BSs was considered in \cite{8403630}. By assuming that the reference aUE is located at a fixed altitude and all the terrestrial BSs are distributed according to a homogeneous Poisson point process (PPP), a tractable coverage probability was derived for aUEs in \cite{8756662}. Further, by assuming that each aUE is equipped with a tilted directional antenna, an exact downlink coverage probability was obtained in \cite{8692749}. On the contrary, for the scenario where a number of terrestrial UEs are served by a single (common) aBS, a joint optimization algorithm was designed to maximize the minimum average rate among multiple UEs in \cite{8685130}. Extending a single aBS to multiple ones, two UAV-based communication scenarios with exact hover time constraints were investigated, and the maximum average data rate was reached for terrestrial UEs in \cite{8053918}. To increase the number of covered terrestrial UEs with minimum transmit power, an optimal placement algorithm regarding aBSs was proposed in \cite{7918510}. Further, an integral expression for the downlink coverage probability of terrestrial UEs was presented in both \cite{7967745} and \cite{8437232}, while a tractable expression on the achievable data rate of terrestrial UEs was given in \cite{8335329}. In \cite{j:SharmaKim2019}, the coverage probability of a terrestrial UE was analytically studied when it is served by multiple aBSs, in the case when the latter nodes make transitions in vertical and spatial directions by following a simple mixed mobility model for $3$D UAV movement process.

Nevertheless, all the aforementioned works were based on the air-to-ground model; either the case when aBSs provide wireless links for terrestrial UEs, or terrestrial BSs offer connectivity for aUEs. The scenario in which aUEs are served by aBSs in an entirely aerial networking platform has not been studied in the appropriate depth so far. In fact, quite recently, a new $3$D cellular architecture for aBSs' deployment based on truncated octahedron shapes was proposed in \cite{8533634}. However, such a cell structure was extremely complex and no closed-form expressions were attained. Besides, concerning UAV-to-UAV systems, the statistical features of the received signal-to-noise ratio (SNR) were defined in \cite{8525328}, while an efficient iterative algorithm to maximize the uplink sum-rate was proposed in \cite{8624565}. To date, both the deployment of $3$D aBSs and performance analysis of a fully-fledged aerial communication network, integrating both aBSs and aUEs, are still open problems.

Further, most of the existing $3$D models for aBSs' deployments are based on Poisson-Voronoi tessellation, which is a direct extension of the $2$D plane \cite{AndrewsTCOM1111, NigamTCOM14s, 8100895}. However, the performance of cell-edge aUEs in Voronoi models is interference limited, as they have farther distance from the serving aBS and experience stronger inter-cell interference (ICI), as compared to the cell-centric aUEs. Coordinated multi-point (CoMP) transmission is a promising method to improve the aUEs' performance, where a reference aUE selects aBSs on a dynamic basis, according to the received strongest signals. In practice, a reference aUE can choose one, two, or more aBSs to cooperate and work with so as to enhance its total received signal strength and overall communication quality. Nonetheless, this approach is achieved at the cost of an overwhelming searching complexity and feedback overhead. 

On another front, most of existing works in the literature related to UAV-based communications are based on PPP modeling, which is in fact not adequate since only a small number of aBSs are required to cover a given finite space. Moreover, although infinite homogeneous PPP has been widely used to model the spatial locations of terrestrial BSs or aBSs, the analytical framework of previous works was essentially relied on the condition that the path loss exponent, say $\alpha$, is greater than the space dimension, which is not suitable for line-of-sight (LoS) communication in the considered $3$D (aerial) free-space platform, where $\alpha \le 3$ usually holds. Particularly, the characteristic functional of a given PPP, say $\Phi$, is not applicable if $\alpha \le 3$ because the total interference $I \triangleq \sum_{x \in \Phi}\|x\|^{-\alpha}$ is almost surely (a.s.) non-convergent, where $\|\cdot\|$ reflects the Euclidean distance. More specifically, by recalling the Campbell's theorem \cite[Theorem~4.6]{Haenggi12}, the total interference is absolutely convergent a.s. if and only if the condition $\int_{\mathbb{R}^3} \min \left(1, \|x\|^{-\alpha} \right) \mathrm{d}x< \infty$ is satisfied \cite[Eq. (4.5)]{Haenggi12}. To investigate this, by applying Campbell's theorem to the mean of total interference, $I$, we have
\begin{align} \label{Eq-1}
	\mathbb{E}(I)& = \mathbb{E}\left(\sum_{x \in \Phi} \|x\|^{-\alpha}  \right) 
			        = \lambda \int_{\mathbb{R}^{3}} \|x\|^{-\alpha} \; \mathrm{d} x \nonumber\\
			     & = \left\{ \begin{array}{rl} \frac{1}{3 - \alpha} 4\pi\lambda r^{3- \alpha} |_0^{\infty}, & \alpha \neq 3; \\
		      		   4\pi\lambda \ln r|_0^{\infty}, & \alpha = 3, \end{array}  \right.     
\end{align}
where $\mathbb{E}[\cdot]$, $r$, and $\lambda$ denote the expectation operator, coverage radius, and density of $\Phi$, respectively. Clearly, if $\alpha > 3$, by properly adapting the path loss model,{\footnote{The divergence of the mean is caused due to the singularity of the path loss law, which is a modeling of artifact, since a receiver cannot get more power than the total transmit power of interfering BSs. To address this issue, the path loss model can be amended as $\ell(x) = \min\{1, \|x\|^{-\alpha}\} ~\mathrm{or}~ \ell(x) = \left(1+ \|x\|^{-\alpha}\right)$. Then, this convergence is guaranteed as $r \rightarrow 0$ and $\mathbb{E}(I)$ is finite, as long as $\alpha > 3$ \cite[p. 146]{Haenggi12}.}} $\mathbb{E}(I)$ is finite and the condition \cite[Eq. (4.5)]{Haenggi12} is satisfied. If $\alpha \le 3$, however, the said condition is not satisfied and the interference computed by \eqref{Eq-1} is infinite.

Unlike PPP, the homogeneous binomial point process (BPP) \cite{Haenggi12, 5299075} is more suitable for UAV-based communications. In fact, BPP has received considerable attention in $2$D cellular and ad-hoc networks \cite{6712183, 6159103}. However, the outage probability was usually derived by assuming that a reference UE was served by a given transmitter (not a point of BPP) at a fixed distance. Recently, considering the scenario that serving BSs for the reference UE will be chosen from the BPP itself, the exact performance analysis for this network was conducted in \cite{7882710}. Using the distance distributions derived in \cite{7882710, 6415342, 5299075}, the downlink coverage probability of terrestrial UEs based on aBSs networks was firstly analyzed  in \cite{7967745}. Nonetheless, BPP has not been used to model $3$D deployments integrating aBSs and aUEs, to the best of our knowledge.

The motivation of this paper is triggered by the recent works \cite{Xia2018Un, 8976426}, where the Poisson-Delaunay triangulation was used to model two-dimensional (2D) cellular networks and a novel CoMP transmission scheme was proposed. This principle is expanded here to model the considered $3$D spatial deployment, which is expected to be an indispensable component of future $5$G-and-beyond space-air-ground integrated networking setups \cite{8368236}. The benefits behind the selection of binomial-Delaunay tetrahedralization used for the newly proposed 3D joint-transmission CoMP (JT-CoMP) are twofold: {\it i)} over all possible tetrahedralizations of a 3D point set, the Delaunay tetrahedralization minimizes the maximum enclosing radius of any simplex, where the enclosing radius of a simplex is defined as the minimum radius of an enclosing sphere \cite{j:VTRajan1994}; and {\it ii)} the cooperation set for an aUE is \emph{fixed and uniquely} determined by the geometric locations of its nearby aBSs. In this study, BPP is used to simulate the aBSs' locations and Delaunay tessellation is used for $3$D cellular network modeling. The resulting network model is called binomial-Delaunay tetrahedralization. This is in contrast to the conventional CoMP in binomial-Voronoi cells, where the said cooperation should be dynamically defined; thereby, causing extensive signaling overhead. Besides, assuming $N$ aBSs and compared to the typical dynamic cooperation in a 3D binomial-Voronoi network, which provides an average (normalized) volume $1/N$ \cite[Table 5.5.2]{b:OkabeBoots2000}, the proposed scheme introduces a corresponding average volume $35/(24 \pi^{2}N) \approx 1/(7N)$ \cite[Table 5.11.2]{b:OkabeBoots2000}. This implies approximately a 7-fold total improvement of the network coverage probability and spectral efficiency.

\subsection{Contributions}
The main contributions of this paper are summarized as follows:
\begin{itemize}
\item[1)] Network model: We propose a novel $3$D cellular model based on binomial-Delaunay tetrahedralization. More specifically, $3$D CoMP transmission is applied (as an extension of the reference $2$D CoMP case in planar terrestrials) in order to further boost the overall communication quality. Particularly, a joint-transmission CoMP (JT-CoMP) scheme is proposed, where each aUE is being simultaneously served by four aBSs using the same system resources, thus forming tetrahedral cells.

\item[2)] Performance analysis: Both the average data rate and coverage probability expressions are derived for the general aUE (i.e., the reference aUE having distinct distances from its four serving aBSs). Corresponding expressions are also provided for the worst-case aUE (i.e., when the reference aUE is located at an equidistant point with regards to its four serving aBSs). 

\item[3)] Frequency planning: A practical frequency allocation is developed so as to effectively mitigate ICI. Based on this strategy, each networking cell operates in different frequency bands according to a fast greedy coloring algorithm. Actually, a practical frequency allocation is proposed, where the most efficient space filling mode, called {\it face-centered cubic} (FCC) packing \cite[Section 6.3]{Conway1999},  is adopted to fill the entire $3$D space. The entire process is indeed a graph coloring problem, where each tetrahedral cell denotes a graph node and the goal is to allocate different colors (i.e., spectrum resources) to each cell appropriately. 
\end{itemize}

To detail the aforementioned contributions, the rest of this paper is organized as follows. Section~\ref{SystemModel} describes the system model and the types of aUEs. Then, Section~\ref{Section-PerformanceAnalysis} is devoted to the performance analysis of both the general aUE and the worst-case aUE, where the achievable data rate and coverage probability of the reference aUE are explicitly derived. Next, a frequency planning scheme is developed in Section~\ref{Section_Frequency}. Simulation and numerical results are discussed in Section~\ref{Section_Simulation} and, finally, Section~\ref{Section_Conclusion} concludes the paper.

\section{System Model}
\label{SystemModel}
\begin{figure}[!t]
	\centering
	\includegraphics[width = 0.35\textheight, clip]{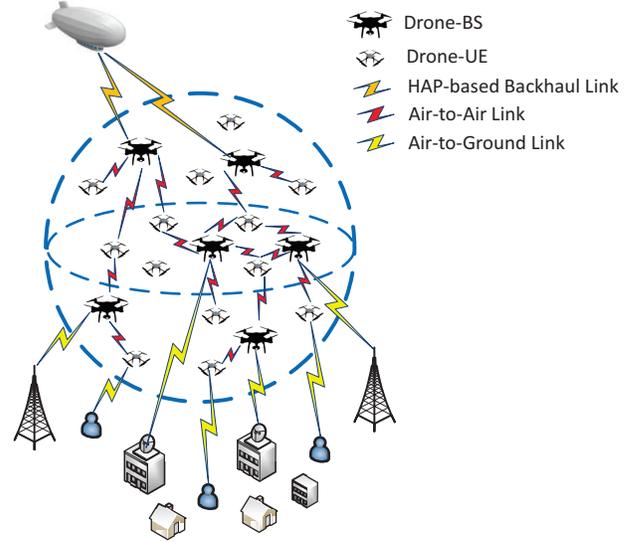}
	\vspace{-10pt}
	\caption{Conceptual model of 3D wireless communication systems.}
	\label{Fig_1}
\end{figure}

\begin{figure} [!t]
	\centering
	\includegraphics [width = 3.75in, clip]{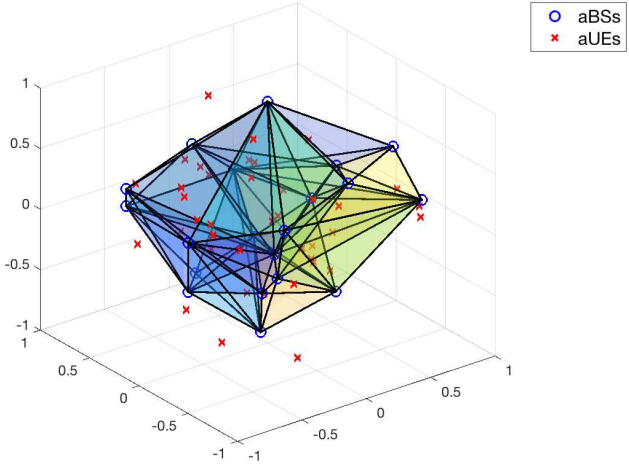}
	\vspace{-20pt}
	\caption{An illustrative cellular network modeled by the binomial-Delaunay tetrahedralization, where the blue-circles refer to the drone aBSs and the cross-marks denote the drone aUEs.}
	\label{Fig_2}
\end{figure}

As illustrated in Fig.~\ref{Fig_1}, we consider a 3D wireless network consisting of UAVs or drones. In particular, let a certain type of drones be aBSs, which serve another type of drone aUEs in the downlink. Further, aBSs are interconnected with the core network via high-altitude platforms (HAPs) and/or terrestrial BSs, while aUEs forward data to their associated terrestrial UEs. Each aBS supports a full 3D connectivity to its corresponding aUEs. Moreover, we consider a BPP network $\Phi$, with $N$ transmitting aBSs uniformly distributed in a finite ball $b(0, R)$ centered at the origin $o = (0, 0, 0)$ with radius~$R$. By using the criteria of the nearest neighbor association, namely, each aUE is directly connected with its nearest aBS, 3D binomial-Voronoi cells are therefore formed. Unfortunately, the typical cell of this Voronoi tessellation is too complex to be mathematically tractable even in $2$D space \cite{Xia2018Un, MonteCarloVoronoi}. It is noteworthy at this point that for an arbitrary dimensional binomial-Voronoi tessellation, its dual graph is the so-called binomial-Delaunay triangulation as shown in Fig.~\ref{Fig_2}. In the $3$D case considered herein, it is known as binomial-Delaunay tetrahedralization. In practice, given the geographical locations of aerial nodes, the binomial-Delaunay tetrahedralization can be uniquely determined by using, e.g., the radial sweep algorithm or divide-and-conquer algorithm \cite[ch.~4]{Hjelle06}. Subsequently, for each aUE, the CoMP cooperation set consists of the four aBS at the vertices of its corresponding tetrahedron.

Like \cite{5299075, 7967745}, a reference aUE is assumed to be located at the origin $(0, 0, 0) \in \mathbb{R}^3$, without loss of generality. By using the propagation model with path loss exponent being $\alpha$, we can express the instantaneous received signal-to-interference-plus-noise ratio (SINR) at the reference aUE as
\begin{equation} \label{Eq_ReceivedSingal}
	\Gamma = \frac{\left| \sum\limits_{i \in \mathcal{C}_0} \kappa_i^{\frac{1}{2}} {P_i^{\frac{1}{2}} d_i^{-\frac{\alpha}{2}}} \right|^2 }{ \sum\limits_{k \in \Phi \setminus \mathcal{C}_0} {\kappa_k P_k d_{k, \, 0}^{-\alpha}} + \sigma_0^2} \; ,
\end{equation}
where the numerator denotes the power of desired signals originating from the cooperating set $\mathcal{C}_0 = \{A_0, B_0, C_0, D_0\}$. The first term in the denominator represents the power of inter-cell interfering signals coming from the remaining aBSs in the set difference $\Phi\setminus \mathcal{C}_0$. Specifically, $\kappa_i$ stands for a fixed (known) system parameter incorporating antenna gains and (reference) propagation attenuation of the considered path loss model. $P_i$ denotes the transmit power of the $i^{\rm th}$ serving aBS while $P_k$ is the transmit power of the $k^{\rm th}$ interfering aBS; $d_i$ indicates the distance between the $i^{\rm th}$ serving aBS and the reference aUE while $d_{k, 0}$ stands for the distance between the $k^{\rm th}$ interfering aBS and the reference aUE; and $\sigma_0^2$ refers to the noise power. Finally, notice that the intra-cell interference is not accounted for in \eqref{Eq_ReceivedSingal} since it can be readily mitigated by classical time and/or frequency division multiplexing techniques. 

Since only a single-tier cellular network and the downlink transmission without power control are considered, the transmit powers of all aBSs are assumed identical and normalized to be unity. Moreover, the fixed system parameters between aBSs and aUEs are assumed identical, that is, $\kappa_i = \kappa_k \triangleq \kappa$, for all $i \in \mathcal{C}_0$ and $k \in \Phi \setminus \mathcal{C}_0$. Meanwhile, as the network performance under study is typically interference-limited, the noise term in \eqref{Eq_ReceivedSingal}, i.e., $\sigma_0^2$, is negligible.\footnote{It is noteworthy that, unlike terrestrial communications, UAV communication links are vulnerable to strong electromagnetic pulse generated by lightning or high‐voltage transmission lines in the complex electromagnetic environment, and the resulting noise must be carefully suppressed \cite{JiaRS20}. After proper noise suppression, the network becomes interference-limited.} Accordingly, by canceling out the constant term $\kappa P$ in \eqref{Eq_ReceivedSingal}, the signal-to-interference ratio (SIR) stems as
\begin{equation} \label{Eq_ReceivedSingal_Model_I}
	\Gamma_1 = \frac{ \left|\sum\limits_{i \in \mathcal{C}_0} d_i^{-\frac{\alpha}{2}} \right|^2}{\sum\limits_{k \in \Phi \setminus \mathcal{C}_0}  { d_{k, \, 0}^{-\alpha}}}.
\end{equation}

\begin{figure}[!t]
	\centering
	\includegraphics [width=1.75in, clip, keepaspectratio]{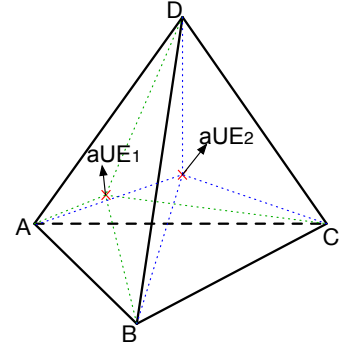}
	\caption{Two types of aUEs: aUE$_1$ has distinct distances to its four serving aBSs $\{A, B, C, D\}$ whereas aUE$_2$ has identical distance to them.}
	\label{Fig_3}
\end{figure}

As illustrated in Fig.~\ref{Fig_3}, we consider two types of aUEs based on their geographical locations relative to aBSs; namely, {\it i)} the general aUE refering to an aUE having distinct distances to its four serving aBSs (i.e., aUE$_1$ in Fig.~\ref{Fig_3}); and {\it ii)} the worst-case aUE which is equidistant from its four serving aBSs (i.e., aUE$_2$ in Fig.~\ref{Fig_3}). The so-called `worst-case' aUE is due to its received signal power is on average smaller than that of any general aUE, by recalling the inequality of arithmetic and geometric means.

To measure the effectiveness of CoMP operation based on the binomial-Delaunay tetrahedralization, we investigate two performance metrics, namely, the achievable data rate and coverage probability. Mathematically, the achievable data rate of a reference aUE can be calculated by
\begin{equation} \label{Definition_rate}
	 \mathcal{R} \triangleq \mathbb{E}\left[\ln(1 + \Gamma)\right],
\end{equation}
where the expectation is taken with respect to the spatial distribution of serving aBSs pertaining to the reference aUE, rather than to channel fading as usual. 

Furthermore, given an outage threshold on the received SIR at a reference aUE, say $\gamma$, the coverage probability is defined as \cite{AndrewsTCOM1111}
\begin{equation} \label{Eq_CoverageProbability}
	\mathcal{P} \triangleq 1 -  \Pr\left\{\Gamma \leq \gamma \right\}.
\end{equation}	

\begin{remark}[On the geometry of the aerial network]
The geometry of the aerial network under study is assumed to be a finite sphere above the ground, for ease of mathematical tractability. However, the BPP based network modeling is also applicable to other possible shapes, such as hemisphere and dome. This is because the BPP needs only the network nodes to be randomly and independently placed in the target space, no matter what the shape of the space is. As for the height of the reference aUE, according to the 3GPP TR 36.777 \cite[Annex A.1]{3GPP36.777}, the height of UAV is set to be uniformly distributed between $1.5$ m and $300$ m. In this regard, the height of the reference aUE can be set to $150$ m.
\end{remark}

\begin{remark}[On the channel model of UAV communications]
	For the ideal scenario in the absence of signal obstruction or reflection, the free-space propagation channel model depends only on the transmitter-receiver distance whereas the effects of shadowing and small-scale fading vanish. This model gives a reasonable approximation when the altitude of UAV is sufficiently high such that a clear LoS link between the UAV and its serving node (either an aerial node or a ground node) is almost guaranteed \cite{8918497}. On the contrary, for low-altitude UAV operating in an urban area where the building height is comparable to UAV altitude, the LoS model is oversimplified and more refined models are necessary. In this paper, we focus on large-scale spatial network modeling and assume that a LoS link exists between any two aerial nodes, thus, we adopt the simplified free-space path loss model, like \cite{8533634, 8525328}. For more details on various channel models of UAV communications, the interested reader is referred to \cite{8918497}.
\end{remark}

\section{Performance  Analysis}
\label{Section-PerformanceAnalysis}
In this section, joint transmission (JT) is applied at the four aBSs within a tetrahedral cooperation set pertaining to both the general aUE and the worst-case aUE, whereas the achievable data rate and coverage probability are used to evaluate their performance.

\subsection{The General aUE}
According to \eqref{Eq_ReceivedSingal_Model_I}, the SIR for the general aUE is explicitly given by
\begin{equation} \label{Eq_ReceivedSingal_Model_2}
	\Gamma_1 = \frac{ \left|\sum\limits_{i =1}^4 d_i^{-\frac{\alpha}{2}} \right|^2}{\sum\limits_{k = 5}^{N} d_{k, \, 0}^{-\alpha} }= \frac{S_1}{I_1},
\end{equation}
where $S_1 \triangleq \left|\sum_{i =1}^4 d_i^{-\frac{\alpha}{2}} \right|^2$ and $I_1 \triangleq \sum_{k = 5}^{N} d_{k, \, 0}^{-\alpha}$. To calculate \eqref{Definition_rate} and \eqref{Eq_CoverageProbability}, we next derive the distance distribution involved in \eqref{Eq_ReceivedSingal_Model_2}.
 Defining the distance between the $i^{\mathrm{th}}$ aBS and the reference aUE at the origin as $D_i$, the cumulative distribution function (CDF) of $D_i$ is given by \cite{5299075}
\begin{equation}\label{CDF_Di}
	F_{D_i}(r_i) =\left( \frac{r_i}{R} \right)^3,~~0\leq r_i \leq R,
\end{equation} 
and, consequently, the probability density function (PDF) of $D_i$ can be readily shown as 
\begin{equation}\label{PDF_Di}
	f_{D_i} (r_i) = \frac{3 r_i^2}{R^3}.
\end{equation}

Conditioned on the serving distance $r$, the set of distances between the reference aUE and the interfering aBSs $\{D_i\}_{i = 5}^N$, are independent and identically distributed (i.i.d.) having the following conditional PDF (for more details, please refer to Appendix~\ref{Proof_Eq_Interference_Distance}):
\begin{equation}\label{PDF_Di_Interfer}
	f_{D_i | r}(r_i) = \frac{3 r_i^2}{R^3 - r^3}, ~~r \leq r_i \leq R.
\end{equation}
Recalling the theory of order statistics \cite[Eq. (2.2.2)]{Shewhart2005Order}, the joint PDF of the nearest four distances is given by
\begin{align}\label{Eq_Joint_2}
	&f_{d_1, d_2, d_3, d_4}\left(r_1, r_2, r_3, r_4\right)   \nonumber\\
	&\quad= \frac{N!}{(N-4)!}\left(1-F_{D_4}(r_4)\right)^{N-4} \prod_{i=1}^{4}f_{D_i}(r_i).
\end{align}
Finally, substituting \eqref{CDF_Di} and \eqref{PDF_Di} into \eqref{Eq_Joint_2} yields
\begin{align} \label{Eq_Joint_Distribution_2}
	&f_{d_1, d_2, d_3, d_4}\left(r_1, r_2, r_3, r_4\right)   \nonumber\\
	&\quad = \frac{N!}{(N-4)!}\left(1-\left(\frac{r_4}{R}\right)^3 \right)^{N-4} \prod_{i=1}^{4}\frac{3r_i^2}{R^3}.
\end{align} 

\subsubsection{Achievable Data Rate}
Now, we are in a position to formalize the achievable data rate of a reference UE in the following theorem. 
	\begin{theorem} \label{Theorem_DataRate_GeneralUEs}
		The achievable data rate of the general aUE can be calculated as
		\begin{align} \label{Theorem_Rate_O_2}
			\mathcal{R}_1 (\alpha) &= \int_{ \bm{r} >0} \int_{z >0} \frac{1}{z}\left(1- M_{S_1}(z)  \right)  M_{I_1}(z) \nonumber\\
			& \quad {}\times f_{d_1, d_2, d_3, d_4}(r_1, r_2, r_3, r_4) \; {\rm d}z \; {\rm d}\bm{r},
		\end{align}
		where $\bm{r} = \left(r_1, r_2, r_3, r_4\right)$ with $r_1 < r_2 < r_3 < r_4$, and the moment generating functions of $S_I$ and $I$ are given by
		\begin{equation}
			M_{S_1}(z) = \exp\left(-z \left|\sum_{i \in \mathcal{C}_0} d_i^{-\frac{\alpha}{2}} \right|^2 \right) \label{Eq_Ms}
		\end{equation} 
		and
		\begin{align}
			M_{I_1}(z) &=  \Big(\frac{3}{\alpha\left(R^{3}-d^{3}_{4}\right)}\left[R^{3}E_{\frac{3+\alpha}{\alpha}}\left(z R^{-\alpha}\right) \right. \nonumber\\
			&\quad \left.\left. {}-d^{3}_{4}E_{\frac{3+\alpha}{\alpha}}\left(z d^{-\alpha}_{4}\right)\right]\right)^{N-4}, \label{Eq_MI_O}
		\end{align}
	respectively, with $E_{v}(\cdot)$ being the $v^{\rm th}$ order exponential integral function \cite[p. xxxiii]{Gradshteyn00}.
	\end{theorem}
\begin{IEEEproof}
	Using the equality involved in the lemma of \cite{Hamdi2010A}, which reads
	\begin{equation}\label{Eq_Equality}
		\ln\left(1+\frac{X}{Y}\right) = \int_{z>0} \frac{1}{z} \left(1-\exp\left(-zX\right)\right)\exp(-zY) \; \mathrm{d}z,
	\end{equation}
	 the achievable data rate can be derived as 
	\begin{align}
		\mathcal{R}_1 (\alpha) &= \int\limits_{\bm{r}>0} \hspace{-0.5em} \mathbb{E}\left[\ln\left(1+\frac{S_1}{I_1}\right)\bigg| \bm{r} \right] f_{d_1, d_2, d_3, d_4}\left(r_1, r_2, r_3, r_4\right) \mathrm{d}\bm{r} \label{Eq_Spectral_a} \\
			&= \int\limits_{\bm{r}>0}f_{d_1, d_2, d_3, d_4}\left(r_1, r_2, r_3, r_4\right) \nonumber\\
			&\quad \times{} \hspace{-0.5em} \int\limits_{z>0}  \frac{1}{z} \left( 1- \underbrace{\mathbb{E}\left[\exp\left(-z S_1\right) \right] }_{M_{S_1}(z)} \right)  \underbrace{\mathbb{E}\left[\exp\left(-z I_1 \right) \right]}_{M_{I_1}(z)} \mathrm{d}z \mathrm{d}\bm{r},\label{Eq_Spectral_b}
	\end{align}
	where $\mathbb{E}\left[\ln\left(1+ {S_1}/{I_1}\right) | \bm{r} \right]$ is the conditional data rate according to \eqref{Definition_rate}, $M_{S_1}(z)$ is previously given by \eqref{Eq_Ms}, and $M_{I_1}(z)$ is derived as follows:
	\begin{align}
		M_{I_1}(z) &= \mathbb{E}_{d_{k, 0}}\left[\exp\left(-z \sum_{k = 5}^{N} d_{k, \, 0}^{-\alpha} \right) \right] \nonumber\\
				&= \mathbb{E}_{d_k} \left[\prod_{k=5}^{N} \exp\left(- z d_{k}^{-\alpha} \right)   \right] \nonumber \\
				&= \left( \mathbb{E}_{d_k}\left[\exp\left(- z d_{k}^{-\alpha} \right)  \right] \right)^{N-4} \nonumber\\
				&= \left( \int_{d_4}^{R} f_{D_i | d_4} (x) \exp\left(-z x^{-\alpha} \right) \mathrm{d}x \right)^{N-4}. \label{Eq_MI_a} 
	\end{align} 
Substituting \eqref{PDF_Di_Interfer} into \eqref{Eq_MI_a} and using \cite[Eq. (8.351.2)]{Gradshteyn00} as well as performing some straightforward manipulations, we get \eqref{Eq_MI_O}. Finally, inserting \eqref{Eq_Ms} and \eqref{Eq_MI_O} into \eqref{Eq_Spectral_b} gives the desired \eqref{Theorem_Rate_O_2}.
	\end{IEEEproof}

\subsubsection{Coverage Probability}
In this subsection, an accurate approximation on the coverage probability for the general aUE is provided. By recalling the causal form of the central limit theorem \cite[p. 234]{Papoulis62}, the sum of multiple positive i.i.d. variables can be approximated by a Gamma distribution. In particular, for the PDF of $I_1$, we have the following lemma.
\begin{lemma} \label{Lemma_PDF_I_1}
	The PDF of $I_1 \triangleq \sum_{k = 5}^{N} d_{k, \, 0}^{-\alpha}$ can be well approximated by the Gamma distribution
		\begin{equation} \label{Eq_Approx_PDF_I_1}
			f_{I_1}(x) \approx \frac{x^{v(d)-1}}{\Gamma[v(d)] \; \theta(d)^{v(d)}} \exp\left( - \frac{x}{\theta(d)}\right),
		\end{equation}
	where
\begin{align}
	v(d) &= \left\{ \begin{array}{rl}
		(N-4)\left[\frac{(3-\alpha)^2\left(R^3 - d^3\right)}{3(3-2\alpha)\left(R^{3-2\alpha} - d^{3-2\alpha}\right) }-1\right]^{-1}, & \alpha \neq 3; \\
		(N-4)\left[\frac{\left(d^{-3} - R^{-3}\right) \left(R^3 - d^3\right)}{9\left(\ln R - \ln d \right)} -1\right]^{-1}, & \alpha = 3,
\end{array}    \right.   \label{Eq_Approx_PDF_I_1_v} \\
\theta(d) &= \left\{ \begin{array}{rl}
			\frac{3-\alpha}{3 - 2\alpha} - \frac{3\left(R^{3- 2\alpha} - d^{3- 2\alpha} \right)}{(3-\alpha)\left(R^3 - d^3   \right)}, & \alpha \neq 3; \\
			\frac{d^{-3} - R^{-3}}{3\left( \ln R -\ln d  \right)} - \frac{3\left( \ln R -\ln d  \right) }{R^3 - d^3}, & \alpha = 3,
			\end{array}     \right. \label{Eq_Approx_PDF_I_1_Theta} 
\end{align}
	and $\Gamma[\cdot]$ denotes the Gamma function \cite[Eq. (8.310.1)]{Gradshteyn00}.
\end{lemma}
\begin{IEEEproof}
	See Appendix~\ref{Proof_Lemma_PDF_I_1}.
\end{IEEEproof}

\begin{figure}[htb]
		\centering
		\includegraphics [width=3.0in, clip, keepaspectratio]{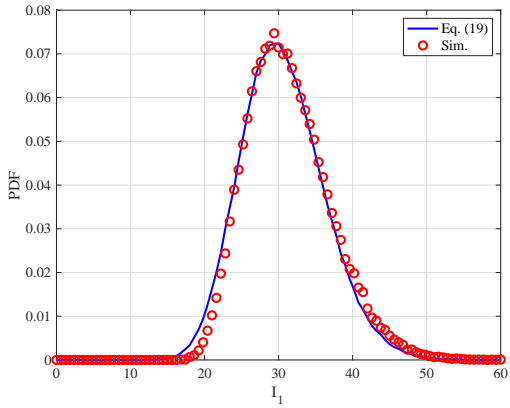}
		\caption{The PDF of the interference $I_1$ ($d = 500$ m, $\alpha = 2.8$ and $N = 150$).}
		\label{Fig-4}
\end{figure}

To illustrate the accuracy of the approximation given by \eqref{Eq_Approx_PDF_I_1}, assuming $d = 500$ m, $\alpha = 2.8$ and $N = 150$, the numerical results computed as per \eqref{Eq_Approx_PDF_I_1} and corresponding simulation ones are compared in Fig.~\ref{Fig-4}. It is clear that they agree well with each other. 

Next, with the aid of Lemma~\ref{Lemma_PDF_I_1}, we can derive an accurate approximation for the coverage probability, as the following theorem suggests.

\begin{theorem} \label{Theorem_Coverage_Finite_General_aUE}
	Given a prescribed outage threshold $\gamma$, the coverage probability of the general aUE can be approximated by 
	\begin{align}\label{Eq_Coverage_1}
			\mathcal{P}_{1}(\gamma, \alpha) &\approx  \frac{1}{\Gamma\left[v(d_4) \right]} \int_{ \bm{r} >0} \gamma \left(v(d_4), \frac{1}{\gamma\, \theta(d_4)} \left|\sum_{i=1}^{4} d_i^{-\frac{\alpha}{2}}\right|^2 \right) \nonumber\\
			&\quad \times{} f_{d_1, d_2, d_3, d_4}(r_1, r_2, r_3, r_4)\, {\rm d}\bm{r},
	\end{align} 
	where $\gamma(\cdot,\cdot)$ is the lower incomplete Gamma function \cite[Eq. (8.350.1)]{Gradshteyn00}, $f_{d_1, d_2, d_3, d_4}(r_1, r_2, r_3, r_4)$ is given by \eqref{Eq_Joint_Distribution_2}, and
$v(d_4)$ and $\theta(d_4)$ are shown in \eqref{Eq_Approx_PDF_I_1_v} and \eqref{Eq_Approx_PDF_I_1_Theta}, respectively. \end{theorem}
\begin{IEEEproof}
	By definition, the coverage probability can be computed as
\begin{equation}\label{Eq_Coverage_1_a}
		\mathcal{P}_{1}(\gamma, \alpha) = \int_{ \bm{r} >0} \mathrm{Pr}\left(\Gamma_1 > \gamma | \bm{r}\right) f_{d_1, d_2, d_3, d_4}(r_1, r_2, r_3, r_4)\, {\rm d}\bm{r},
\end{equation}
where the conditional probability in the integrand becomes 
\begin{align}
		\mathrm{Pr}\left(\Gamma_1 > \gamma | \bm{r} \right) &= \mathrm{Pr} \left(I_1 < \frac{1}{\gamma} \left|\sum_{i=1}^{4} d_i^{-\frac{\alpha}{2}}\right|^2  \right) \label{CDF_Gamma_1_(a)}\\
		& \approx  \frac{1}{\Gamma\left[v(d_4)\right]} \gamma \left(v(d_4), \frac{1}{\gamma\, \theta(d_4)} \left|\sum_{i=1}^{4} d_i^{-\frac{\alpha}{2}}\right|^2 \right).\label{CDF_Gamma_1}
\end{align}
As earlier stated, $I_1$ is the sum of multiple conditional i.i.d. random variables. Therefore, by applying Lemma~\ref{Lemma_PDF_I_1}, \eqref{CDF_Gamma_1_(a)} can be explicitly computed as \eqref{CDF_Gamma_1} via \cite[Eq. (3.381.1)]{Gradshteyn00}. Finally, substituting \eqref{CDF_Gamma_1} and the joint distance distribution \eqref{Eq_Joint_Distribution_2} into \eqref{Eq_Coverage_1_a} yields the intended \eqref{Eq_Coverage_1}.
\end{IEEEproof}

Now, we establish a connection between Theorems~\ref{Theorem_DataRate_GeneralUEs} and~\ref{Theorem_Coverage_Finite_General_aUE}. Specifically, as per \cite[Eq. (5)]{9013645}, given the coverage probability $\mathcal{P}_{1}(\gamma, \alpha)$, the achievable data rate can be computed as
\begin{align} \label{Eq_ConnectionThem1and2}
	\mathcal{R}_1 (\alpha) &\triangleq \mathbb{E}[\ln(1+\Gamma_1)] = \int_{0}^{\infty} \ln(1+ \gamma) f_{\Gamma_1} (\gamma) \; \mathrm{d} \gamma \nonumber\\
	&\overset{(a)}{=}\int_{0}^{\infty} \frac{1}{1+\omega} \left[\int_{\omega}^{\infty} f_{\Gamma_1} (\gamma) \; \mathrm{d}\gamma\right] \mathrm{d} \omega \nonumber\\
	&= \int_{0}^{\infty} \frac{\mathcal{P}_1(\gamma, \alpha)}{1+\gamma} \; \mathrm{d} \gamma,
\end{align}
where $f_{\Gamma_1} (\gamma)$ denotes the PDF of $\Gamma_1$ and the integral $\ln(1+ \gamma) = \int_{0}^{\gamma} {\frac{1}{1+\omega}} \mathrm{d} \omega$ is exploited in the step $(a)$. As to be shown at the end of Section~\ref{Sim_General_aUE}, the numerical results computed by \eqref{Eq_ConnectionThem1and2} coincide with those computed as per \eqref{Theorem_Rate_O_2}, which cross-validates the preceding derivations. 

\subsection{The Worst-Case aUE}
\label{Section_WorstCaseUE_2}
In this subsection, the achievable data rate and coverage probability are analyzed for the worst-case aUE. By recalling Fig.~\ref{Fig_3}, an aUE at the vertex of a tetrahedral cell, i.e., ${\rm aUE}_2$, is chosen as the reference point and set to be the origin in 3D space, which implies that the Euclidean distances between the reference aUE and its serving aBSs are identical, i.e., $d_i = d$, for all $i \in \mathcal{C}_0$. Thereby, \eqref{Eq_ReceivedSingal_Model_I} reduces to
\begin{equation} \label{Eq_ReceivedSingal_WorstCase_2}
\Gamma_2 = \frac{ 16 d^{-\alpha}}{\sum\limits_{k = 5}^{N} d_{k, \, 0}^{-\alpha} }.
\end{equation}
Also, the PDF of the equidistant $d$ can be derived and given by (for more details, please refer to Appendix~\ref{Proof_Eq_PDF_EqualDistance})
\begin{align}\label{Eq_PDF_EqualDistance}
	f_{d}(x)=\frac{3}{R}\frac{1}{\beta\left(N-3, 3 \right)} \left(\frac{x}{R}\right)^{8}\left(1-\left(\frac{x}{R}\right)^3 \right)^{N-4}.
\end{align} 

\subsubsection{Achievable Data Rate}
Based on \eqref{Eq_Equality} and \eqref{Eq_PDF_EqualDistance}, the achievable data rate of the worst-case aUE is formalized in the following theorem.
\begin{theorem} \label{Theorem_Rate_Worstcase_1'}
	The achievable data rate of the worst-case aUE can be computed as
	\begin{equation} \label{Eq_Rate_Vertex_2}
		\mathcal{R}_2(\alpha) =  \int_{0<x<R} \int_{z >0} \frac{1}{z}\left(1- M_{S_2}(z)  \right)  M_{I_2}(z) \, f_{d}(x) \; {\rm d}z \; {\rm d}x,
	\end{equation}
	where 
	\begin{align}
    	M_{S_2}(z) &= \exp\left(-16 z d^{-\alpha} \right), \nonumber\\
		M_{I_2}(z) &= \Big(\frac{3}{\alpha\left(R^{3}-d^{3}\right)}\left[R^{3}E_{\frac{3+\alpha}{\alpha}}\left(z R^{-\alpha}\right) \right. \nonumber\\
	   &\quad \left.\left. {}-d^{3} E_{\frac{3+\alpha}{\alpha}}\left(z d^{-\alpha} \right)\right]\right)^{N-4}.
	\end{align}
\end{theorem}
\begin{IEEEproof}
	The proof is similar to that of Theorem~\ref{Theorem_DataRate_GeneralUEs} and, thus, it is omitted for brevity.
\end{IEEEproof}

\subsubsection{Coverage Probability}
Using the same method as in Theorem~\ref{Theorem_Coverage_Finite_General_aUE}, the coverage probability for the worst-case aUE can be derived, as summarized in the following theorem.
\begin{theorem} \label{Theorem_Coverage_Finite_Worst_aUE}
	The coverage probability of the worst-case aUE is approximated by 
	\begin{equation} \label{Eq_Coverage_2'}
		\mathcal{P}_{2} (\gamma, \alpha) \approx  \frac{1}{\Gamma\left[v(d) \right]} \int_{ d>0} \gamma \left(v(d), \frac{16 d^{-\alpha}}{\gamma\, \theta(d)} \right) f_{d}(x)\, {\rm d}x,
	\end{equation} 
where $f_d(x)$ is given by \eqref{Eq_PDF_EqualDistance}, and $v(d)$ and $\theta(d)$ are shown in \eqref{Eq_Approx_PDF_I_1_v} and \eqref{Eq_Approx_PDF_I_1_Theta}, respectively. 
	\end{theorem}
	
Likewise, \eqref{Eq_ConnectionThem1and2} can be used to make a connection between \eqref{Eq_Rate_Vertex_2} and \eqref{Eq_Coverage_2'}.

\section{Frequency Planning and Interference Mitigation}
\label{Section_Frequency}
This section develops a practical frequency planning strategy so as to sufficiently reduce ICI. In particular, the frequency division multiple access (FDMA) scheme is adopted herein.\footnote{Other orthogonal multiple access schemes can alternatively be used instead, such as time division multiple access (TDMA) or orthogonal FDMA (OFDMA), without affecting the main results presented hereinafter.} Doing so, each four-node aBS set forming a tetrahedral cell occupies a unique frequency band for its underlying aUEs. Nonetheless, a given aBS could be interconnected simultaneously to multiple tetrahedra, reflecting that each aBS should be able to reserve a corresponding number of available (i.e., non-overlapping, non-occupied) frequency bands. Therefore, the proposed drone-based wireless cellular network can be visualized as a 3D colored graph, consisting of multiple non-overlapping tetrahedra. Each tetrahedron has a unique color reflecting the corresponding frequency band to serve its associated aUEs. Also, the number of distinct frequency bands that a given aBS must preserve is determined by its corresponding graph degree and the number of aUEs inside that every tetrahedron is linked with. For a sufficient spectrum resource, whenever the number of aBSs is relatively low (e.g., $N < 20$), the said condition may be feasible. However, as the number of aBSs increases, the corresponding spectrum resources per each cell become quite narrow. To this end, a frequency reuse scheme is required in accordance to the typical 2D cellular infrastructures.

\subsection{Face-Centered Cubic Packing}
\begin{figure}[t!] 
	\centering    
	\subfloat[Close-packing of equal spheres of radius $r$.] 
	{
		\begin{minipage}[t]{0.5\textwidth}
			\centering         
			\includegraphics[width=0.5\textwidth]{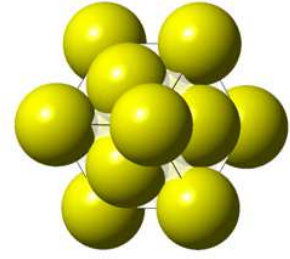}   
			\label{Fig_5a}
		\end{minipage}
	}
	
	\subfloat[FCC packing in a cubic of edge length $a$.] 
	{
		\begin{minipage}[t]{0.5\textwidth}
			\centering      
			\includegraphics[width= 0.5\textwidth]{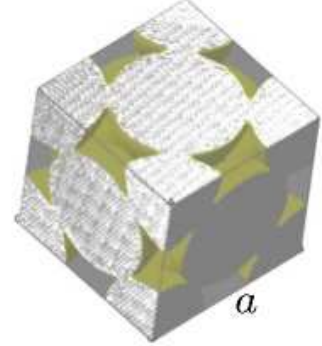}  
			\label{Fig_5b} 
		\end{minipage}
	}
	\caption{The principle of face-centered cubic (FCC) packing.} 
	\label{Fig-5}  
\end{figure}

As illustrated in Fig.~\ref{Fig-5}, the {\it face-centered cubic} (FCC) packing is widely adopted to fill the entire $3$D space \cite[Section 6.3]{Conway1999}. In particular, Fig.~\ref{Fig_5a} shows the close-packing of equal spheres of radius $r$, which is in geometry a dense arrangement of congruent spheres in an infinite, regular manner (i.e., lattice). Figure~\ref{Fig_5b} illustrates FCC packing in a cubic of edge length $a$, where the spheres are situated at the corners of the cubic and at the centers of all the cube faces. To guarantee a close-packed arrangement, namely,  there is no way to pack more spheres into the cubic, we must have $a = 2\sqrt{2}r$ and, hence, the packing efficiency is $\delta = \mathcal{V}_{\mathrm{Sphere}}/\mathcal{V}_{\mathrm{Cube}} = (16\pi r^3/3)/(16\sqrt{2}r^3) \simeq 74\%$. As proved in \cite{Hales05}, this is the highest packing efficiency achieved by a lattice packing. 

Now, we integrate the methodology of FCC packing into our binomial-Delaunay tetrahedralization for frequency planning. The main idea is to use each sphere in Fig.~\ref{Fig_5a} to represent a cluster of tetrahedra. The same spectrum resource is reused among spheres whereas in each sphere, the spectrum resource is orthogonally allocated to different tetrahedra. As a result,  the frequency allocation strategy is a solution to the well-known graph coloring problem, where each tetrahedron denotes a graph node and the goal is to allocate different colors (i.e., frequency bands) to each tetrahedron lying within its corresponding sphere. 

\begin{figure}[!t]
	\centering
	\includegraphics [width=3.5in, clip, keepaspectratio]{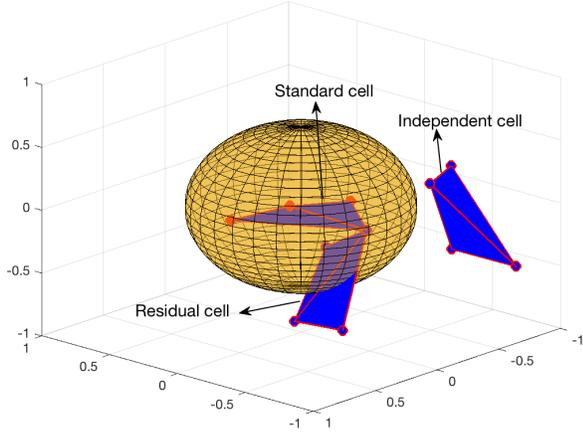}
	\vspace{-25pt}
	\caption{Three types of tetrahedral cells with respect to a reference sphere, where `$\circ$' denotes the drone aBSs.}
	\label{Fig-6}  
\end{figure}

\begin{figure}[!t]
	\centering
	\includegraphics [width=2.0in, clip, keepaspectratio]{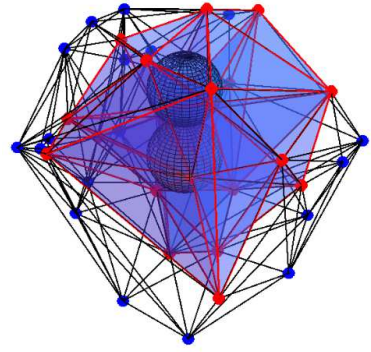}
	\vspace{-10pt}
	\caption{An illustrative example of CoMP transmission based on binomial-Delaunay tetrahedralization with two filling spheres, where red circles denote the drone aBSs involved in the two observed spheres while the blue circles refer to the remaining aBSs.}
	\label{Fig-7}
\end{figure}

Prior to further proceeding, as shown in Fig.~\ref{Fig-6}, we define three types of tetrahedral cells: {\it i)} \emph{Standard cell}: It refers to a tetrahedron whose vertices (thus, its whole volume) are located inside a given sphere; {\it ii)} \emph{Residual cell}: It stands for a tetrahedron intersecting a sphere. In other words, the volume of a residual cell does not entirely belong to a particular sphere and it may be `spatially connected' to more than one sphere; and {\it iii)} \emph{Independent cell}: It indicates a tetrahedron whose volume falls entirely outside any given sphere, i.e., in the gap between tangent spheres. For illustration purposes, Fig.~\ref{Fig-7} depicts two filling spheres (one is blue and the other is yellow), and the red circles refer to aBSs involved in the two spheres whereas the blue circles refer to the remaining aBSs. The tetrahedra with red boundaries illustrates the three types of cells involved in the two spheres. 

\subsection{Frequency Reuse Distance}
To further elaborate on the frequency reuse criterion, the \emph{effective interference radius}, $\epsilon$, is defined such that each interfering aBS using the same frequency band should be outside this region. The achievable data rate is selected as a standard performance metric. In particular, the following condition should be satisfied:
\begin{align} \label{condRate}
	\mathbb{E}\left[\ln(1 + \Gamma)\right]\geq \mathcal{R}_{\rm th},
\end{align}   
where $\mathcal{R}_{\rm th}$ is a predetermined minimum data rate of every aUE in the network under study (in nat/sec/Hz). Since we consider a fully-fledged aerial wireless cellular system consisting entirely of drones, the path loss exponent is set to $\alpha=2$. Also, let $b(o, \epsilon)$ be a sphere with radius $\epsilon$ centered at the origin $o$, $X\triangleq |\sum_{i \in \mathcal{C}_0} d_i^{-1} |^2$, and $Y_{\epsilon}\triangleq \sum_{k \in \Phi \setminus b(o, \epsilon)}  { d_{k, \, 0}^{-2}}$. It is straightforward that
\begin{align}
	\mathbb{E}\left[\ln\left(1 + \frac{X}{Y_{\epsilon}}\right)\right] 
		\leq \ln\left(1 + \mathbb{E}\left[\frac{X}{Y_{\epsilon}}\right]\right) 
		\approx \ln\left(1 + \frac{\mathbb{E}\left[X\right]}{\mathbb{E}\left[Y_{\epsilon}\right]}\right),
\label{condRate1}
\end{align} 
which implies that a large $\epsilon$ is desired from the received SIR viewpoint as $\mathbb{E}\left[Y_{\epsilon}\right]$ decreases with $\epsilon$. In contrast, as the said sphere $b(o,\epsilon)$ gets larger, there are more tetrahedral cells in a cluster and each cell gets less spectrum resources. To resolve this dilemma, the optimal spherical radius, $\epsilon^{\star}$, is determined by allowing the maximum interference, that is, the inequality \eqref{condRate} takes equality: 
\begin{align} \label{condRate2}
	\ln\left(1 + \frac{\mathbb{E}\left[X\right]}{\mathbb{E}\left[Y_{\epsilon^{\star}}\right]}\right)=\mathcal{R}_{\rm th},
\end{align} 
where $\mathbb{E}\left[Y_{\epsilon^{\star}}\right]$ can be computed as
\begin{align} \label{meanInt}
	\mathbb{E}\left[Y_{\epsilon^{\star}}\right] 
		&= N \frac{\epsilon^{\star 3}}{R^3} \int_{\epsilon^{\star}}^{R}x^{-2} \frac{3 x^2}{R^3-\epsilon^{\star 3}} \; \mathrm{d} x \nonumber \\
		&= \frac{3N\epsilon^{\star 3}}{R^3\left(R^{2} + R\epsilon^{\star} + \epsilon^{\star 2}\right)},
\end{align} 
and $\mathbb{E}[X]$ in \eqref{condRate2} is evaluated by
\begin{align} 
	\mathbb{E}[X] &= \int_{\bm{r}>0}\left(r^{-1}_{1} + r^{-1}_{2} + r^{-1}_{3} + r^{-1}_{4}\right)^{2} \nonumber\\
			      &\quad {}\times f_{d_{1}, d_{2}, d_{3}, d_{4}}(r_{1}, r_{2}, r_{3}, r_{4}) \; \mathrm{d}\bm{r} \label{expXeval} \\
			      &= \frac{65 \Gamma(N+1) \Gamma\left(\frac{10}{3}\right)}{12 \epsilon^{2} \Gamma\left(N+\frac{1}{3}\right)}, \label{expXevalclosedform}
\end{align}
with $\{\bm{r}: 0<r_{1}\leq r_{2}\leq r_{3}\leq r_{4}\leq \epsilon\}$. As for the worst-case aUE, we have
\begin{align} \label{expXevalclosedform2}
	\mathbb{E}[X] 
		= 16\int^{\epsilon}_{0} r^{-2}f_{d}(r) \; \mathrm{d}r 
		= \frac{8 \Gamma\left(\frac{7}{3}\right) \Gamma\left(N\right)}{\epsilon^{2}\Gamma\left(N-\frac{2}{3}\right)},
\end{align}
where $f_{d}(\cdot)$ is defined earlier in \eqref{Eq_PDF_EqualDistance}. 

Next, inserting \eqref{expXevalclosedform} (resp., \eqref{expXevalclosedform2}) and \eqref{meanInt} into \eqref{condRate2} for the general aUEs (resp., the worst-case aUEs), the numerical result for the optimal spherical radius $\epsilon^{\star}$ can be computed as per
\begin{align} \label{Eq_epsilon}
	\left(e^{\mathcal{R}^{\mathrm{th}}}-1 \right) \epsilon^{\star 5} - \Delta(N) R^3 \epsilon^{\star 2} - \Delta(N) R^4 \epsilon^{\star} - \Delta(N) R^5 = 0,
\end{align}
with $\Delta(N) \triangleq 65 \Gamma(N+1) \Gamma\left(10/3\right)/\left(36 N\Gamma\left(N+1/3\right)\right)$ for the general aUE, whereas for the worst-case aUE $\Delta(N) \triangleq 8 \Gamma(N) \Gamma\left(7/3\right)/\left(3 N\Gamma\left(N- 2/3\right)\right)$. The solution $\epsilon^{\star}$ for the worst-case aUE provides an upper bound for the general aUE.

Based on the definition of the sphere $b(o,\epsilon^{\star})$, a cluster of manifold tetrahedral cells is being formed lying within this sphere. Each cell in a given cluster uses distinct spectrum resources so as to eliminate intra-cluster interference. Outside $b(o,\epsilon^{\star})$, the same spectrum resources are reused by other cells forming another cluster, and the entire process is quite similar to the classical planar case. Therefore, frequency reuse factor can be defined as
\begin{align} \label{freqreusefactDefinition}
	\eta \triangleq \left\lceil \frac{\mathcal{V}_{b(o,\epsilon^{\star})}}{\mathbb{E}\left[\mathcal{V}_{\rm cell}\right]} \right\rceil, 
\end{align}
where the numerator and denominator represents the volume of the effective interfering sphere and the expected volume of the reference cell, respectively, and $\left\lceil \cdot \right\rceil$ stands for the ceiling operator. The volume of the sphere $b(o,\epsilon^{\star})$ is simply $\mathcal{V}_{b(o,\epsilon^{\star})} = 4 \pi (\epsilon^{\star})^{3}/3$. Further, according to \cite[Table 5.11.2]{b:OkabeBoots2000} and \cite[Thm. 2.9 and Def. 2.12]{Haenggi12}, we get $\mathbb{E}\left[\mathcal{V}_{\rm cell}\right]=35 R^{3}/(18 \pi N)$. As a result, the integer-valued frequency reuse factor can be explicitly computed as
\begin{align} \label{freqreusefactDefinition1}
	\eta \triangleq \left\lceil \frac{24}{35}N \pi^2 \left(\frac{ \epsilon^{\star}}{R}\right)^{3} \right\rceil.
\end{align}

\subsection{Frequency Allocation}
The basic idea of the proposed frequency allocation strategy is as follows. Given a spherical space of radius $R$, there are $N$ aBSs with each predetermined data rate threshold $\mathcal{R}_{\rm th}$. We first construct the Delaunay tetrahedralization and define the corresponding tetrahedral cells by using, e.g., the radial sweep algorithm or divide-and-conquer algorithm \cite[Ch.~4]{Hjelle06}. Then, we compute the radius of sphere $\epsilon^{\star}$ as per \eqref{Eq_epsilon}, and fill the target space with spheres of radius $\epsilon^{\star}$ by using FCC packing. For each sphere, count its standard cells and residual cells and, then, sort the resulting spheres in a descending order by starting with the one with the largest number of tetrahedral cells, which is labeled Sphere-1. Given that there are $k_1$ cells in Sphere-1, the bandwidth reserved for each cell is simply $\mathcal{B}/k_1$, where $\mathcal{B}$ denotes the total bandwidth for the whole system. Finally, apply a fast greedy coloring algorithm \cite{graphcoloring} with repeated random sequences to assign different spectrum resources for each cell, starting with Sphere-$1$. This colored pattern is repeated in every sphere such that unique colors are being assigned to cells of the same sphere, meanwhile, the effective interference radius $\epsilon^{\star}$ is always satisfied for each cell. Finally,  all independent cells are identified and randomly assigned colors chosen from Sphere-1.

\begin{figure}[!t]
	\centering
	\includegraphics[width=3.0in, clip, keepaspectratio]{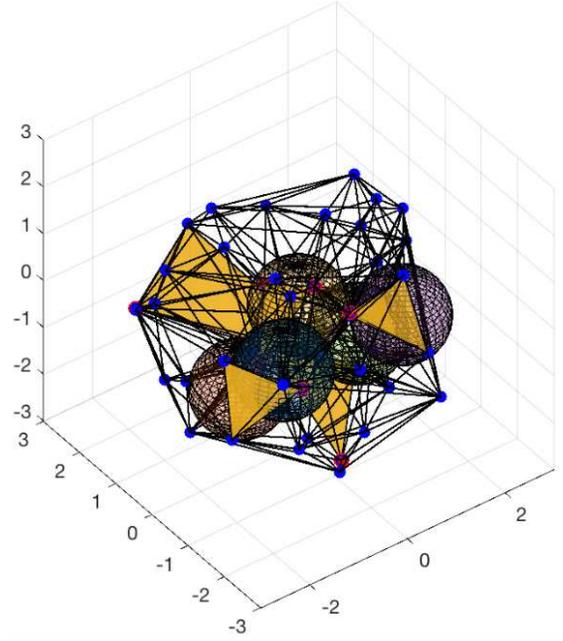}
	\caption{An illustrative example of the proposed frequency allocation, where the five yellow tetrahedral cells pertaining to five distinct spheres are assigned the same frequency resource.}
	\label{Fig-8}
\end{figure}

In summary, the proposed frequency allocation based on greedy coloring is formalized in Algorithm~\ref{alg_DuplicateRemoval}. For illustration purposes, Fig.~\ref{Fig-8} shows five spheres and five tetrahedral cells of the same frequency band, filled in yellow.

\begin{algorithm} [!t]
	\caption{Greedy-Coloring based Frequency Allocation}
	\label{alg_DuplicateRemoval}
	\begin{algorithmic}[1]  
		\REQUIRE The radius $R$ of a 3D space, the number of aBSs $N$, and the threshold of data rate $\mathcal{R}_{\rm th}$.
		\STATE {Construct the Delaunay tetrahedralization;}
		\STATE {Calculate the optimal spherical radius, $\epsilon^{\star}$, as per \eqref{Eq_epsilon};}
		\STATE {Fill the space with spheres of radius $\epsilon^{\star}$ by using FCC packing;}
		\STATE {For each sphere, count the number of its standard cells and residual cells;}
		\STATE {Sort the spheres in a descending order by starting with the one having the largest number of cells (say, $k_1$), and define $W = \{k_1, k_2, \cdots, k_n\}$, where $k_1 \ge k_2 \ge \cdots \ge k_n$, where $n$ denotes the number of spheres;}
		\FOR {$i := 1$ to $n$ \do} 
			\IF {Sphere-$i$ has partially colored tetrahedral cells}
				\STATE {Assign the smallest number of possible colors to the remaining tetrahedral cells;}
			\ENDIF
		\ENDFOR 
		\STATE {Identify independent cells;}
		\STATE {Randomly assign colors (chosen from Sphere-$1$) to independent cells.}
	\end{algorithmic}
\end{algorithm}

\begin{remark}[The maximum number of colors]
	In general, the optimal approach for frequency planning is to adopt a greedy coloring algorithm. However, greedy coloring is a well-known NP-hard problem. For practical applications, a heuristic solution is adopted in this paper. Obviously, Algorithm~\ref{alg_DuplicateRemoval} is suboptimal yet practical. The reason is that the color allocation of the underlying $3$D graph is unbalanced since the maximum number of different colors (i.e., $k_1$) is determined by the most dense sphere in a single round. On the other hand, the optimal solution would require multiple iterations of the used coloring algorithm (e.g., balanced coloring \cite{graphcoloring}) to determine a potentially smaller number of colors. Nevertheless, the latter gain in terms of spectrum resources could be obtained via a considerably higher computational complexity. 
\end{remark}

\subsection{Interference Analysis}
When frequency reuse is implemented amongst the available tetrahedral cells, the interfering aBSs, not all aBSs in $\Phi$, which transmit in the same frequency band, are a thinned version of the original BPP with a total number of aBSs $N' \triangleq (N-4)/\eta$. Since a thinned version of a BPP is again a BPP, the achievable data rate for general aUEs can be computed as
\begin{align}
	\mathcal{R}_1' (\alpha, \eta) &= \frac{1}{\eta}\int_{ \bm{r} >0} \int_{z >0} \frac{1}{z}\left(1- M_{S_1}(z)  \right)  M_{I_1}'\left(z, \eta \right) \nonumber\\
						   & \quad {}\times f_{d_1, d_2, d_3, d_4}(r_1, r_2, r_3, r_4) \; {\rm d}z \; {\rm d}\bm{r},\label{Rate_Thinned_Reuse}
\end{align}
where $\bm{r} = \left(r_1, r_2, r_3, r_4\right)$ with $r_1 < r_2 < r_3 < r_4$, and $M_{S_1}(z) $ is given by \eqref{Eq_Ms}, and 
\begin{align}
	M_{I_1}'(z, \eta) &= \left( \frac{1}{R^3 - d_4^3} \int_{d_4}^{R} \hspace{-0.5em} 3x^2 \exp\left(-z x^{-\alpha} \right) \mathrm{d}x \right)^{\frac{N-4}{\eta}} \\
				&= \left(\frac{3}{\alpha\left(R^{3}-d^{3}_{4}\right)}\left[R^{3}E_{\frac{3+\alpha}{\alpha}}\left(z R^{-\alpha}\right)\right.\right. \nonumber\\
				&\quad \left.\left. {}-d^{3}_{4}E_{\frac{3+\alpha}{\alpha}}\left(z d^{-\alpha}_{4}\right)\right]\right)^{\frac{N -4}{\eta}}\label{Eq_MI_O_Reuse}.
\end{align}
The same analogy holds also for the worst-case scenario. 

Apart from the above thinning BPP, the arising point process (regarding the inter-cluster interference) can be sufficiently modeled by a Mat\'{e}rn hard-core point process (MHCPP) of Type~I \cite{Matern}; whereby, there are no points inside $b(o,\epsilon^{\star})$, while they are uniformly placed elsewhere. Given a parent PPP with intensity~$\lambda$, Type I MHCPP has a corresponding intensity $\lambda'\triangleq \lambda \exp\left(-\lambda 4 \pi (\epsilon^{\star})^{3}/3\right)$ \cite[Section 3.5]{Haenggi12}. To evaluate the interfering power at the cluster-based system model, the approximation in \cite{5934671} is adopted where the Type I MHCPP is approximated as a PPP, which follows the conditional intensity $\lambda'$. Doing so, referring back to our model with a given number of aBSs $N$ and with the help of \cite[Thm. 2.9 and Def. 2.12]{Haenggi12}, the inter-cluster interference is modeled by a BPP with parameter 
\begin{align}
	\psi \triangleq \left\lfloor (N-4) \exp\left(-(N-4) \frac{{\epsilon^{\star}}^{3}}{R^3}\right)\right\rceil,
	\label{psiParameter}
	\end{align}
	where $\left\lfloor \cdot \right\rceil$ is the rounding operator to the closest integer. Then, the  equivalent frequency reuse factor is 
	\begin{align}
	\eta' = \left\lfloor \frac{N-4}{\psi} \right\rceil. \label{DefinitionEta2}
\end{align}
	Consequently, the achievable data rate of the general aUE can be expressed as
\begin{align}
	\mathcal{R}_1' (\alpha, \eta') &= \frac{1}{\eta'}\int_{ \bm{r} >0} \int_{z >0} \frac{1}{z}\left(1- M_{S_1}(z)  \right)  M_{I_1}'\left(z, \eta'\right) \nonumber\\
	&\quad {}\times f_{d_{1}, d_{2}, d_{3}, d_{4}}(r_1, r_2, r_3, r_4) \; {\rm d}z \; {\rm d}\bm{r}, \label{Rate_HardCore_Reuse}
\end{align}
where $M_{S_1}(z)$ and $M_{I_1}'(z, \eta')$ are given by \eqref{Eq_Ms} and \eqref{Eq_MI_O_Reuse}, respectively.

\section{Simulation Results and Discussions}
\label{Section_Simulation}
In this section, numerical results computed as per the previously obtained analytical expressions are illustrated, in comparison with extensive Monte-Carlo simulation results.  In the pertaining simulation experiments, a large 3D wireless network with coverage radius $R = 3$ km is assumed. Since we focus on LoS signal propagation conditions in free-space, the value of path loss exponent is set to $\alpha \leq 3.2$.\footnote{Although the path loss exponent $\alpha \le 3.2$ is  assumed in our simulation experiments, it can take values $\alpha > 3.2$. More specifically, the value of $\alpha$ is environment-related and ranges approximately from $1.6$ (e.g., hallways inside buildings) to $8$ (e.g., dense urban environments)\cite[p. 41]{Goldsmith05}, with $\alpha = 2$ corresponding to free-space propagation.}

\subsection{Achievable Data Rate and Coverage Probability}

\subsubsection{General aUEs}
\label{Sim_General_aUE}

\begin{figure}[!t]
	\centering
	\includegraphics [width=3.2in, clip, keepaspectratio]{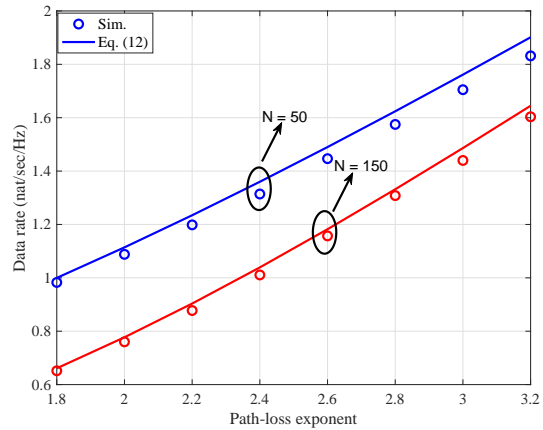}
	\caption{Achievable data rate of the general aUEs versus the path loss exponent $\alpha$ with $N = 50, \ 150$.}
	\label{Fig-9}
\end{figure}

Figure~\ref{Fig-9} depicts the achievable data rate versus the path loss exponent $\alpha$ for general aUEs, where the number of aBSs is set to $N =50$ or $N =150$. As observed, for a fixed $N$, the data rate monotonically increases with $\alpha$. It implies that the interfering power decreases more intense compared to the signal power, as $\alpha$ increases; thereby, improving the achievable data rate. On the other hand, for a fixed $\alpha$, the achievable data rate decreases for a larger number of aBSs, as interference gets stronger with regards to an increased number of interfering aBSs. For either case, the numerical results computed as per Theorem~\ref{Theorem_DataRate_GeneralUEs} agree well with the simulation ones, which corroborates the effectiveness of our analysis. 

\begin{figure}[!t]
	\centering
	\includegraphics [width=3.2in, clip, keepaspectratio]{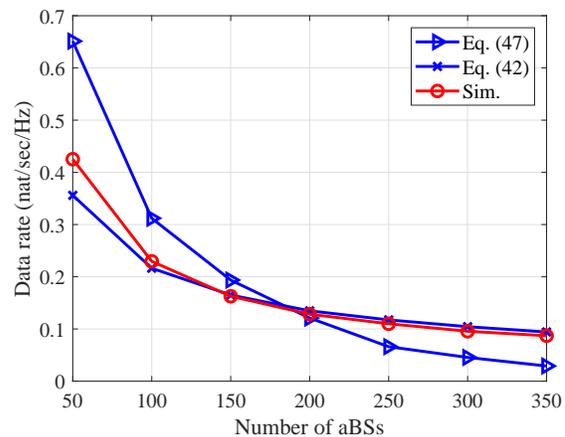}
	\caption{Achievable data rate of the general aUEs versus the number of aBSs, $\alpha =2$ and $\mathcal{R}_{\mathrm{th}} = \ln(1+10^3)\approx 6.9$ nats/sec/Hz.}
	\label{Fig-10}  
\end{figure}

Figure~\ref{Fig-10} shows the achievable data rate versus different number of aBSs, $N$, for a given threshold data rate $\mathcal{R}_{\mathrm{th}} = \ln \left(1 + 10^3 \right)\approx 6.9$ nats/sec/Hz. The corresponding frequency reuse factor is calculated via \eqref{freqreusefactDefinition1} for the random interference model and \eqref{DefinitionEta2} for the Type I MHCPP interference model. In fact, the average data rate decreases with an increased density of aBSs, since the frequency reuse factor is increasing as the number of aBSs also increases for a fixed data rate threshold, while the data rate decreases with an increased reuse factor. On the other hand, it can be seen that the numerical results of \eqref{Rate_Thinned_Reuse} based on a thinning BPP are sharply close to the simulated ones, compared to the ones based on \eqref{Rate_HardCore_Reuse}. This is consistent to our approach, adopting the tractable randomized frequency assignment scheme. It is also revealed that the correlation between points in the Type I MHCPP make the properties of a BPP no longer valid.

\begin{figure}[!t]
	\centering
	\includegraphics[width=3.2in, clip]{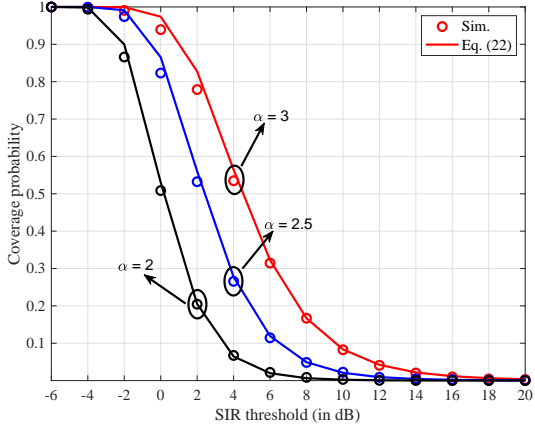}
	\caption{Coverage probability of the general aUEs versus the SIR threshold in the unit of dB, with $N = 150$.}
	\label{Fig-11}
\end{figure}

\begin{figure}[!t]
	\centering
	\includegraphics [width=3.2in, clip, keepaspectratio]{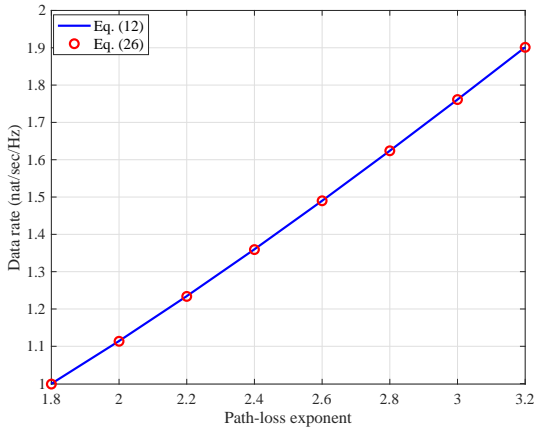}
	\caption{Achievable data rate of the general aUEs.}
	\label{Fig-12}
\end{figure}

Figure~\ref{Fig-11} illustrates the coverage probability versus the SIR threshold, with different path loss exponent~$\alpha$. It is seen that the coverage probability increases with $\alpha$, as larger $\alpha$ implies lower interference. Besides, the numerical results computed as per Theorem~\ref{Theorem_Coverage_Finite_General_aUE} also match well with the corresponding simulated ones, which further verifies the accuracy of the proposed approach. Finally, Fig.~\ref{Fig-12} illustrates the achievable data rates numerically computed by \eqref{Theorem_Rate_O_2} and \eqref{Eq_ConnectionThem1and2}. Their complete coincidence further validates Theorems~1 and 2.

\subsubsection{Worst-case aUEs}

\begin{figure}[!t]
	\centering
	\includegraphics [width=3.2in, clip, keepaspectratio]{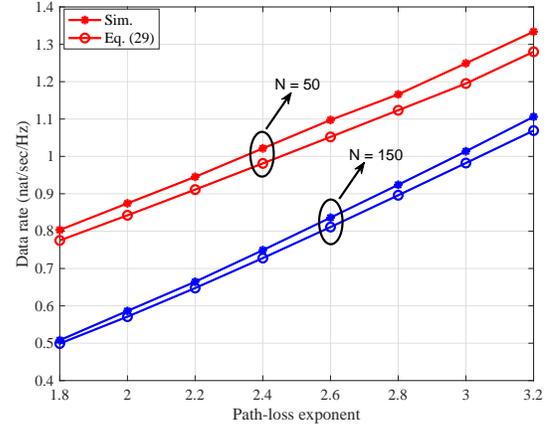}
	\caption{Achievable data rate of the worst-case aUEs versus the path loss exponent $\alpha$ with  $N = 50$ and $N = 150$.}
	\label{Fig-13}
\end{figure}

\begin{figure}[t]
	\centering
	\includegraphics[width=3.2in, clip]{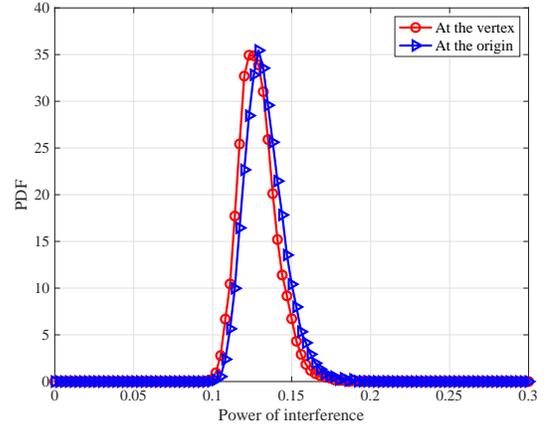}
	\caption{PDFs of the interference at the origin and at the vertex.}
	\label{Fig-14}
\end{figure}

\begin{figure}[!t]
	\centering
	\includegraphics [width=3.2in, clip, keepaspectratio]{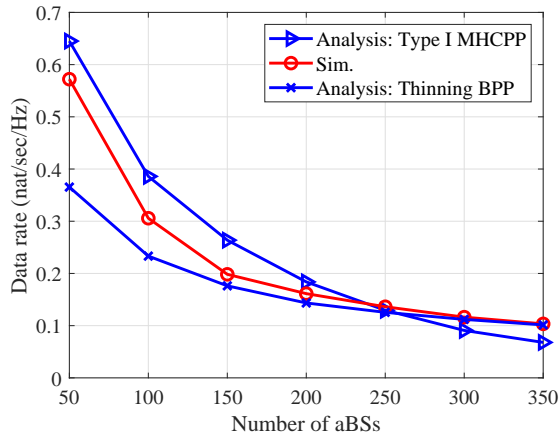}
	\caption{Achievable data rate of the worst-case aUEs versus the number of aBSs $N$ with $\alpha =2$ and $\mathcal{R}_{\mathrm{th}} = \ln(1+10^3)\approx 6.9$ nats/sec/Hz.}
	\label{Fig-15}  
\end{figure}

\begin{figure}[t]
	\centering
	\includegraphics[width=3.2in, clip]{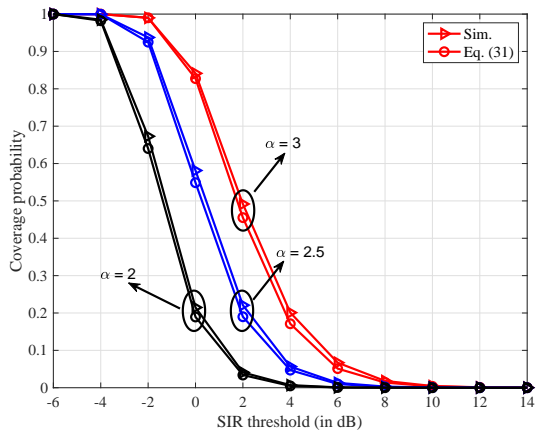}
	\caption{Coverage probabilities of the worst-case aUEs versus the SIR threshold in the unit of dB with $N = 150$.}
	\label{Fig-16}
\end{figure}

Figure~\ref{Fig-13} shows the achievable data rate versus the path loss exponent $\alpha$ for the worst-case aUEs. It is noted that the analytical results lower bound the simulation ones. In fact, this underestimation is introduced by the assumption of independence between aBSs and a reference aUE. More specifically, by recalling the Slivnyak-Mecke theorem in stochastic geometry \cite[p. 132]{Chiu13}, a reference aUE is assumed to be located at the origin $(0, 0, 0) \in \mathbb{R}^3$, without loss of generality. This assumption implies that the location of a reference aUE is independent of aBSs' location. However, as far as the worst-case aUE is concerned, a reference aUE maintains the same distance from its four nearest aBSs. Obviously, the location of a reference aUE here directly depends on the locations of its closest aBSs. Such a dependence results in an overestimation of the total interfering power as further shown in Fig.~\ref{Fig-14}. The same observation is found in Fig.~\ref{Fig-15}, where the analytical results are a bit lower than the simulation ones. It is noteworthy that, although the achievable data rate decreases with more and more aBSs, the simulated achievable data rate of the worst-case aUEs is higher than that of the general aUEs as shown in Fig.~\ref{Fig-10}, which serves as a performance upper bound. The same observation is also reflected in Fig.~\ref{Fig-16}, where the analytical results slightly deviate (i.e., overestimate the total interfering power) from the corresponding simulated ones. It is remarkable that, although the interfering power is approximated by the Gamma distribution, the said deviation caused by the interference approximation given by \eqref{Eq_Approx_PDF_I_1} is rather marginal (the analytical coverage probability is well matched with the simulation results for the general aUE case as shown in Fig.~\ref{Fig-11}). The reason for such a slight deviation is the independence assumption of the reference aUE.

\subsection{Binomial-Delaunay Tetrahedralization vs. Binomial-Voronoi Tessellation}

\subsubsection{Comparison with binomial-Voronoi tessellation without CoMP} 

\begin{figure}[t]
	\centering
	\includegraphics[width=3.2in, clip]{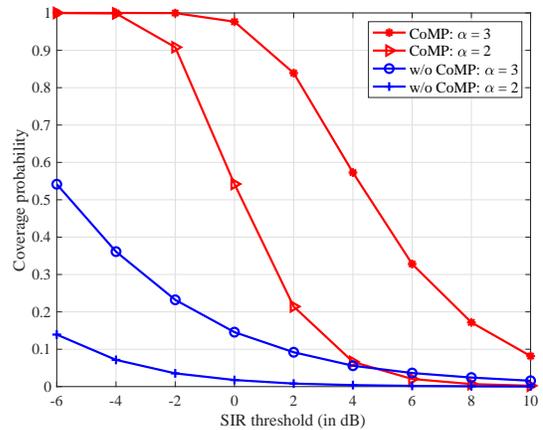}
	\caption{Coverage probabilities of the general aUEs versus the SIR threshold in the unit of dB under the proposed Delaunay CoMP scheme and binomial-Voronoi tessellation scheme without CoMP and $N =150$.}
	\label{Fig-17}
\end{figure}

Figure~\ref{Fig-17} illustrates the coverage probability of the general aUEs versus the SIR threshold, where the path loss exponent was set to $\alpha =2$ or $\alpha =3$. It is observed that the coverage probability in our proposed model is much higher than that in the scheme without CoMP, even at a relatively high value of $\alpha = 3$. This emphatically manifests that cooperative communication is quite necessary, especially in the presence of ICI.

\subsubsection{Comparison with binomial-Voronoi tessellation with dynamic cooperation set}

\begin{figure}[t]
	\centering
	\includegraphics[width=3.2in, clip]{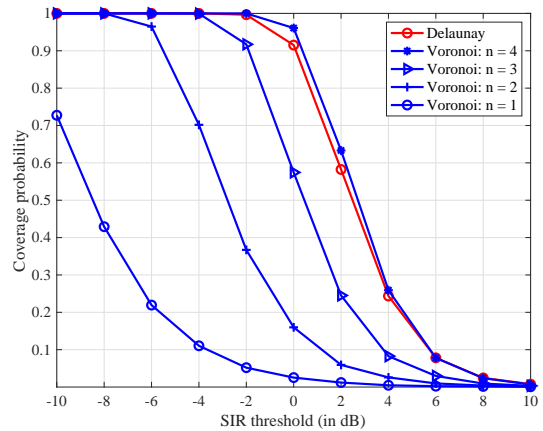}
	\caption{Coverage probabilities for a reference aUE simulated under the proposed Delaunay CoMP scheme and binomial-Voronoi tessellation scheme with $n = \{1, 2, 3, 4\}$ dynamic cooperating BSs and $N =50$.}
	\label{Fig-18}
\end{figure}

For binomial-Voronoi tessellation utilizing a dynamic cooperation set, the reference aUE is always located at the center of its serving aBSs, belonging to a aUE-centric scheme. In this scheme, the reference aUE exhaustively searches every aBS so as to determine the appropriate aBS set whereby it receives the strongest signals. Then, aUE can choose one, two, or more aBSs to cooperate and enhance its received SINR, at the cost of searching complexity and feedback (signaling) overhead. It turns out that the resultant aBS set does not always include the four nearest aBSs. This scheme is fundamentally different from our proposed approach, which is aBS-centric (i.e., all aUEs within a cell are being served by an identical fixed and location-dependent aBS set). For the worst-case aUE, the serving aBSs are the four nearest aBSs with equal distance. On this condition, the signal power in the aUE-centric case is equivalent to that of an aBS-centric case; the same outcome also holds for the interfering power. For the general aUE case, without loss of generality, we choose the origin as the reference aUE. The simulation results are presented in Fig.~\ref{Fig-18}. It can be seen that the performance under the two schemes is comparable, which further implies that the serving aBSs at the vertices of a tetrahedral cell in our approach can be well approximated as the nearest four ones. Most importantly, it is verified that the proposed approach presents quite a similar performance as the conventional approach; yet, introducing considerably reduced computational efforts and feedback overhead.

\section{Conclusion}
\label{Section_Conclusion}
In this paper, a novel $3$D cellular model was proposed. In particular, CoMP was adopted to enhance the communication quality, based on the binomial-Delaunay tetrahedralization; whereby, each aUE in a tetrahedral cell can be jointly served by four aBSs, thus providing reliable and high throughput connectivity. Analytical formulae regarding the achievable data rate and coverage probability were offered for two user cases, namely, the general aUE and the worst-case aUE. It turns out that CoMP brings significant performance gain to the considered system setup in the presence of ICI. Also, a fair comparison with the classical binomial-Voronoi approach with a dynamic cooperation set was conducted. Simulation and numerical results explicitly indicated that our proposed approach is comparable to the dynamic scheme, yet introducing much less computational burden and signaling overhead. Finally, a practical frequency allocation scheme based on a fast greedy coloring algorithm was developed, which provides a benchmark for future network planning in $3$D aerial networking infrastructures.

\appendices
\numberwithin{equation}{section}
\section{Derivation of Eq.~\eqref{PDF_Di_Interfer}}
\label{Proof_Eq_Interference_Distance}
Given the serving distance $d_4$, the joint conditional PDF of the ordered distances ${d_i}$, $i = 5:N$, is given by
	\begin{equation}\label{Proof_PDF_JointDistance}
		f_{d_5, d_6, \cdots, d_N | d_4}\left(r_5, r_6, \cdots, r_N \right) = \frac{f_{d_4, d_5, \cdots, d_N}\left(r_4, r_5, \cdots, r_N \right)}{f_{d_4}\left(r_4\right)},
	\end{equation}
	where the numerator can be explicitly computed as
	\begin{align}
		&f_{d_4, d_5, \cdots, d_N}\left(r_4, r_5, \cdots, r_N \right) \nonumber\\
		&= \int\limits_{r_1>0} \int\limits_{r_2 > r_1} \int\limits_{r_3 > r_2}  f_{d_1, d_2, \cdots, d_N}\left(r_1, r_2, \cdots, r_N \right) \mathrm{d}r_3 \; \mathrm{d}r_2 \; \mathrm{d}r_1 \nonumber\\
		&=\int\limits_{r_1>0} \int\limits_{r_2 > r_1} \int\limits_{r_3 > r_2} N! \prod_{i=1}^{N} f_{D_i} \left(r_i\right) \mathrm{d}r_3 \; \mathrm{d}r_2 \; \mathrm{d}r_1 \label{PDF_joint_a}\\
		&= N! \prod_{i=4}^{N} f_{D_i} \left(r_i\right) \int\limits_{r_1>0} \int\limits_{r_2 > r_1} \int\limits_{r_3 > r_2}  \prod_{i=1}^{3} f_{D_i} \left(r_i\right) \mathrm{d}r_3 \; \mathrm{d}r_2 \; \mathrm{d}r_1 \nonumber\\
		&=N! \prod_{i=4}^{N} f_{D_i} \left(r_i\right) \frac{1}{3!}\prod_{i=1}^{3}\int_{0}^{r_3} f_{D_i}\left(r_i\right)  \mathrm{d}r_i \label{PDF_joint_b}\\
		&=\frac{N!}{3!} \prod_{i=4}^{N} f_{D_i} \left(r_i\right) \left[F_{D}(r_3)\right]^3,    \label{PDF_joint_c}
	\end{align}
	where \eqref{PDF_joint_a} follows from the definition of the joint distance distribution based on the equality in \cite[Eq. (2.10)]{Ahsanullah2005Order}, and \eqref{PDF_joint_b} follows from the symmetry of \cite[Eq. (2.12)]{Ahsanullah2005Order}. According to  \cite[Eq. (4)]{5299075}, the PDF of the fourth nearest distance can be shown as  
\begin{equation} \label{Eq_PDF_d4}
	f_{d_4}(r_4) = \frac{N!}{3!(N-4)!}\left[F_D\left(r_4\right)\right]^3 \left[1-F_D\left(r_4\right)\right]^{N-4}f_{D}\left(r_4\right).
\end{equation}
Inserting \eqref{PDF_joint_c} and \eqref{Eq_PDF_d4} into \eqref{Proof_PDF_JointDistance} yields
\begin{equation}
	f_{d_5, d_6, \cdots, d_N | d_4}\left(r_5, r_6, \cdots, r_N \right) = (N-4)!\prod_{i=5}^{N} \frac{f_{D}\left(r_i\right)}{1-F_D\left(r_4\right)}.
\end{equation}
Note that there are $(N-4)!$ possible permutations of the ordered distances $d_i$, $i = 5, 6, \cdots, N$. Since the interfering aBSs are chosen uniformly at random, the permutation term does not appear in the conditional joint PDF of the unordered distances \cite[Appendix~A]{7882710}, that is,
\begin{equation}
	f_{d_5, d_6, \cdots, d_N | d_4}\left(r_5, r_6, \cdots, r_N \right) = \prod_{i=5}^{N} \frac{f_{D}\left(r_i\right)}{1-F_D\left(r_4\right)}, 
\end{equation}
then, the sampling distribution of the $N-4$ i.i.d. random variables is given by 
\begin{equation}\label{Eq_PDF_Interference_Distance}
f_{D_i | r} (r_i) = \frac{f_{D}\left(r_i\right)}{1-F_D\left(r_4\right)},
\end{equation}
where $f_D\left(r_i\right)$ is shown in \eqref{PDF_Di}. Finally, substituting \eqref{CDF_Di}  and \eqref{PDF_Di} into \eqref{Eq_PDF_Interference_Distance} yields the intended \eqref{PDF_Di_Interfer}.

\section{Proof of Lemma~\ref{Lemma_PDF_I_1}}
\label{Proof_Lemma_PDF_I_1}
By recalling the causal form of the central limit theorem \cite[p. 234]{Papoulis62}, it is known that the distribution of the sum $I_1 \triangleq \sum_{k = 5}^{N} d_{k, \, 0}^{-\alpha}$ can be approximated by the Gamma distribution given by \eqref{Eq_Approx_PDF_I_1}, with the parameters determined by
\begin{equation}\label{Eq_v}
		v(d) = \frac{\mathbb{E}^2(I_1)}{\mathrm{Var}(I_1)}, \quad \theta(d) = \frac{\mathrm{Var}(I_1)}{\mathbb{E}(I_1)},
\end{equation}
where the operators $\mathbb{E}(I_1)$ and $\mathrm{Var}(I_1)$ denote the mean and variance of $I_1$, respectively.

By virtue of \eqref{PDF_Di_Interfer}, the average interference power conditioned on the serving distance $d_4$ is computed as
\begin{align}\label{Eq_mean}
	\mathbb{E}(I_1) &= (N-4) \int_{d}^{R}x^{-\alpha} \frac{3 x^2}{R^3-d^3} \; \mathrm{d} x \nonumber\\
							&= \left\{ \begin{array}{rl}
								\frac{3(N-4)}{(3-\alpha)(R^3- d^3)} \left(R^{3-\alpha} - d^{3-\alpha}\right), & \alpha \neq 3, \\ 
								\frac{3(N-4)}{R^3- d^3}\left(\ln{R} - \ln{d} \right), & \alpha = 3.
								\end{array} \right. 
\end{align}
Likewise, the conditional variance of interference power can be derived as 
\begin{align}
	&\mathrm{Var}(I_1) = (N-4)\left[\int\limits_{d}^{R} \frac{3 x^{2-2\alpha}}{R^3- d^3} \mathrm{d} x - \left(\int_{d}^{R} \frac{3x^{2-\alpha}}{R^3 - d^3} \mathrm{d}x   \right)^2  \right] \nonumber \\
								&= \left\{ \begin{array}{rl}
								\hspace{-0.5em} (N-4) \left[\frac{3(R^{3-2\alpha} -d^{3-2\alpha})}{(3-2\alpha)(R^3 - d^3)} - \frac{9\left(R^{3-2\alpha} -d^{3-2\alpha} \right)^2}{(3-\alpha)^2\left(R^3 - d^3\right)^2} \right], & \hspace{-0.5em} \alpha \neq 3, \\ 
								\hspace{-0.5em} (N-4)\left[\frac{d^{-3} - R^{-3}}{R^3 - d^3} - \left(\frac{3}{R^3 - d^3} \left(\ln R -\ln d \right)  \right)^2  \right], & \hspace{-0.5em} \alpha =3.
								\end{array}  \right. \label{Eq_variance}
\end{align}
Finally, in light of \eqref{Eq_v}-\eqref{Eq_variance} and after performing some algebraic manipulations, we attain the desired \eqref{Eq_Approx_PDF_I_1_v} and \eqref{Eq_Approx_PDF_I_1_Theta}.

\begin{figure*}[t]
	\begin{align*} \label{Eq_F_d}
	F_{d}{(\delta)} = \lim\limits_{\epsilon \rightarrow 0} \frac{\int_{0}^{\delta} \mathrm{Pr}\left[\Phi \left(b(o,x)\right) = 0, \Phi \left(b(o, x+\epsilon) \setminus b(o, x) = k \right) | \Phi(W) = N\right] \mathrm{d}x }
		{\int_{0}^{R} \mathrm{Pr}\left[\Phi \left(b(o,x)\right) = 0, \Phi \left(b(o, x+\epsilon) \setminus b(o, x) = k \right), \Phi(W) = N \right] \mathrm{d}x}.
	\tag{C.2} 
	\end{align*}
\end{figure*}

\section{Derivation of Eq.~\eqref{Eq_PDF_EqualDistance}}
\label{Proof_Eq_PDF_EqualDistance}
Define a new process $\Phi(W) =N$, i.e., a process having exactly $N$ points uniformly distributed in $W$. The resulting set is a binomial point process, which is in fact a conditional PPP $\Phi$ with $\Phi(W) =N$ \cite[p. 43]{Chiu13}. Let $b(o, r)$ be a $3$D ball of radius $r$ centered at $o$. The conditional probability that there are $k$ points in $b(o, r)$ is given by \cite[Eq. (13)]{Moltchanov2012Distance}:
\begin{align}
&\mathrm{Pr}\left[\Phi[b(o, r) ] =k | \Phi (W) = N\right] \nonumber\\
	&\qquad = \binom{N}{k} p^k (1-p)^{N-k}, \ 0 \leq r \leq R ,\label{Eq_Probability_}
\end{align}
where the operator $\binom{\cdot}{\cdot}$ denotes the binomial coefficient, $p = {V\left[ b(o, r) \cap W \right] }/{V[W]} = (r/R)^{3}$ with $V\left[b(o, r) \cap W \right]$ and $V[W]$ being the volumes of $b(o, r) \cap W$ and $W$, respectively.

Denoting the distance between the vertex and its nearest $k$ neighbors by $d$, in principle, its CDF $F_{d}(\delta)$ can be calculated as per \eqref{Eq_F_d} shown at the top of the next page. For further proceeding, by using a similar technique as that in \cite[Sec.~6]{Muche2005The}, \eqref{Eq_F_d} can be simplified as
 \begin{align}
 	F_{d}(\delta) &= \lim\limits_{\epsilon \rightarrow 0} \frac{\int\limits_{0}^{\delta} \binom{N}{k} \left[\frac{1}{R^{3}}\left( (x+\epsilon)^{3}-x^{3} \right) \right]^k \left[ 1-\left(\frac{x}{R}\right)^{3} \right]^{N-k} \hspace{-1em} \mathrm{d}x }
		{\int\limits_{0}^{R} \binom{N}{k} \left[\frac{1}{R^{3}}\left( (x+\epsilon)^{3}-x^{3} \right) \right]^k \left[1-\left(\frac{x}{R}\right)^{3} \right]^{N-k}  \hspace{-1em} \mathrm{d}x } \nonumber\\
 			&= \frac{\int\limits_{0}^{\delta} x^{(3-1)k} \left[ 1-\left(\frac{x}{R}\right)^{3} \right]^{N-k} \mathrm{d}x } {\int\limits_{0}^{R} x^{(3-1)k} \left[ 1-\left(\frac{x}{R}\right)^{3} \right]^{N-k} \mathrm{d}x}. \tag{C.3}
 \end{align}
Then, taking differentiation of $F_{d}(\delta)$ with respect to $\delta$ and using \cite[Eq. (8.380.1)]{Gradshteyn00}, we attain
\begin{equation}\label{Proof_Worst_Distance_PDF}
	f_{d}(\delta) = \frac{3}{R\beta\left( N-k, \left(\frac{2}{3}\right)^k+\frac{1}{3} \right)} \left(\frac{\delta}{R}\right)^{2k}\left[1-\left(\frac{\delta}{R}\right)^{3} \right]^{N-k}, \tag{C.4}
\end{equation}
where $\beta(x, y) \triangleq \int_{0}^{1}t^{x-1} (1-t)^{y-1} \mathrm{d}t$ is the Beta function. Finally, substituting $k=4$ into \eqref{Proof_Worst_Distance_PDF} yields the desired \eqref{Eq_PDF_EqualDistance}.

\bibliographystyle{IEEEtran}
\bibliography{References}

\begin{IEEEbiography}
	[{\includegraphics[width=1in, height=1.25in, clip, keepaspectratio]{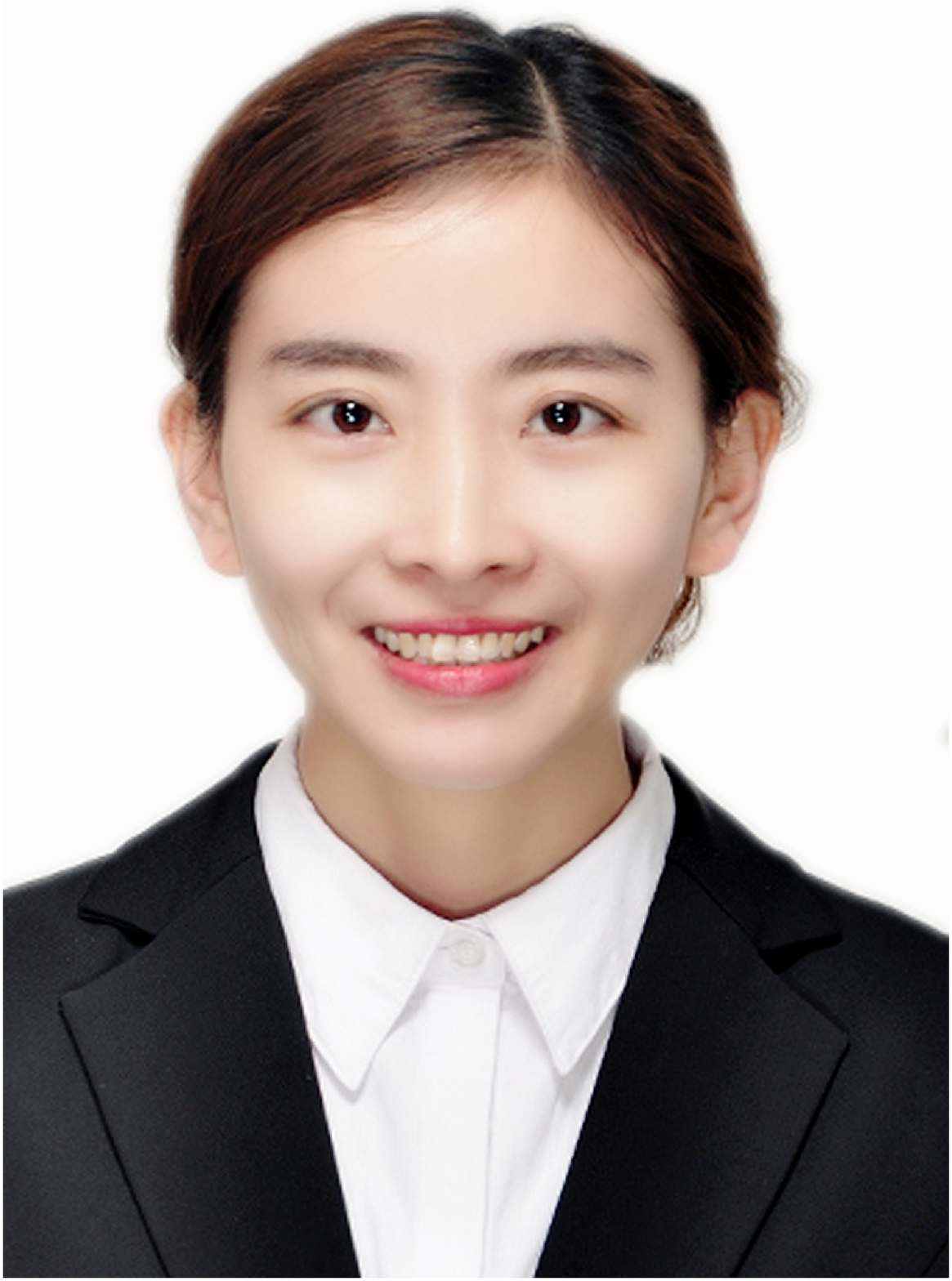}}]{Yan Li}  received the B.S. degree in Electronic Information Engineering from Hunan Normal University, Changsha, China, in 2013, and the M.S. degree in Electronics and Communication Engineering from Sun Yat-sen University, Guangzhou, China, in 2016. She is currently working towards the PH.D. degree in Information and Communication Engineering at Sun Yat-sen University. Her research interests include modeling and analysis of cellular networks based on stochastic geometry theory, cooperative communications.
\end{IEEEbiography}

\begin{IEEEbiography}
	[{\includegraphics[width=1in, height=1.25in, clip, keepaspectratio]{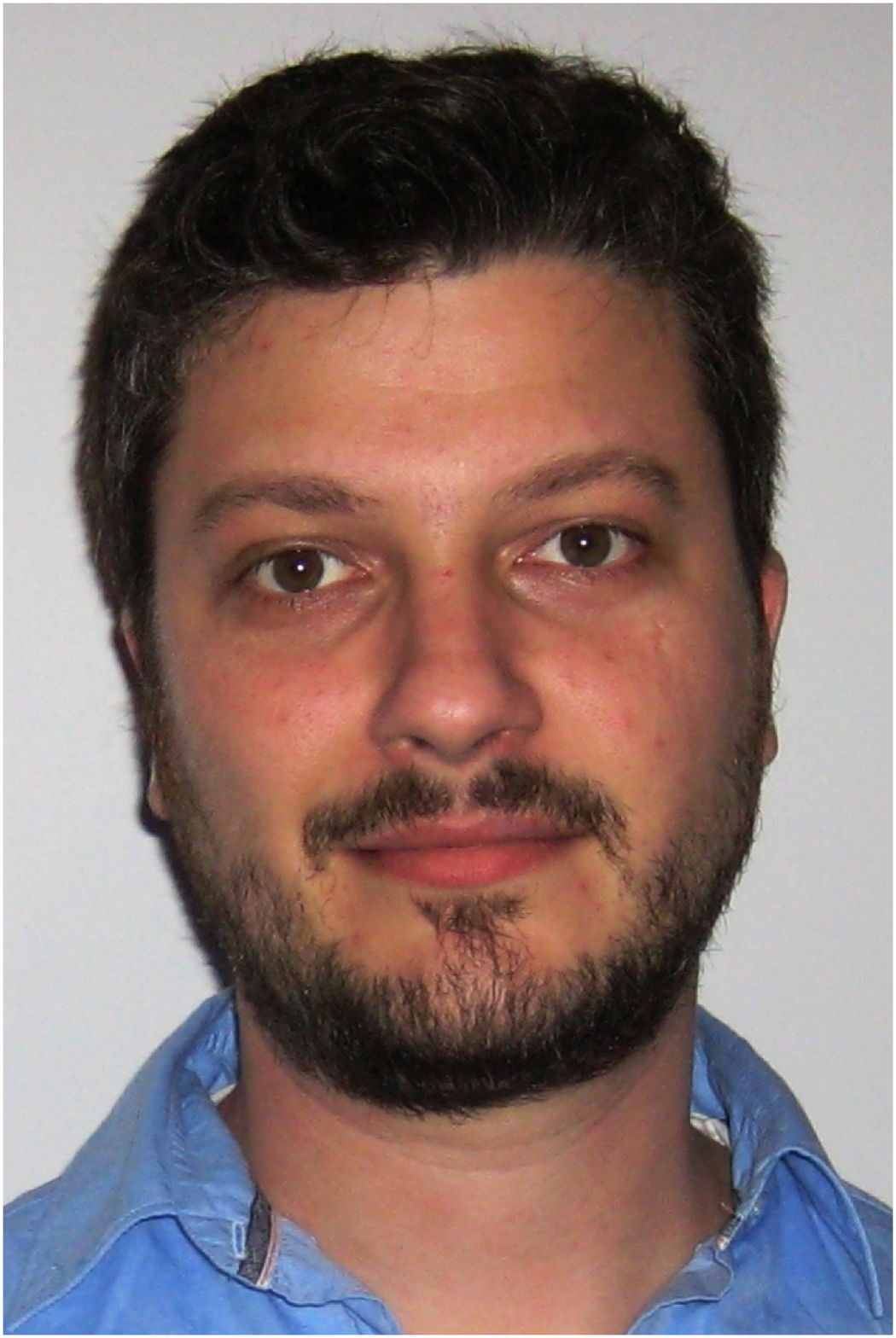}}]{Nikolaos I. Miridakis} (M'18, SM'19) was born in Athens, Greece, in 1982. He received his M.Sc.  and Ph.D. degrees in Networking and Data Communications from the Department of Information Systems, Kingston University, U.K. in 2008 and from the Department of Informatics, University of Piraeus, Greece in 2012, respectively. Since 2012, he has been with the Department of Informatics, University of Piraeus, Greece where he is a Senior Researcher. Also, since 2018, he has been with the School of Electrical and Information Engineering and the Institute of Physical Internet, Jinan University, Zhuhai, China as a Distinguished Research Associate. Currently, he is an Assistant Professor with the Department of Informatics and Computer Engineering, University of West Attica, Greece. His main research interests include wireless communications, and more specifically interference analysis and management in wireless communications, multicarrier communications, MIMO systems, statistical signal processing, diversity reception, fading channels, and cooperative communications.

   Dr. Miridakis serves as a reviewer and TPC member for several prestigious international journals and conferences. He was also recognized as an Exemplary Reviewer by \textsc{IEEE Transactions on Communications}, \textsc{IEEE Transactions on Vehicular Technology}, and \textit{Elsevier Physical Communication} in 2017. Since 2019, he serves as an Associate Editor of the \textsc{IEEE Communications Letters}.
\end{IEEEbiography}

\begin{IEEEbiography}
	[{\includegraphics[width=1in, height=1.25in, clip, keepaspectratio]{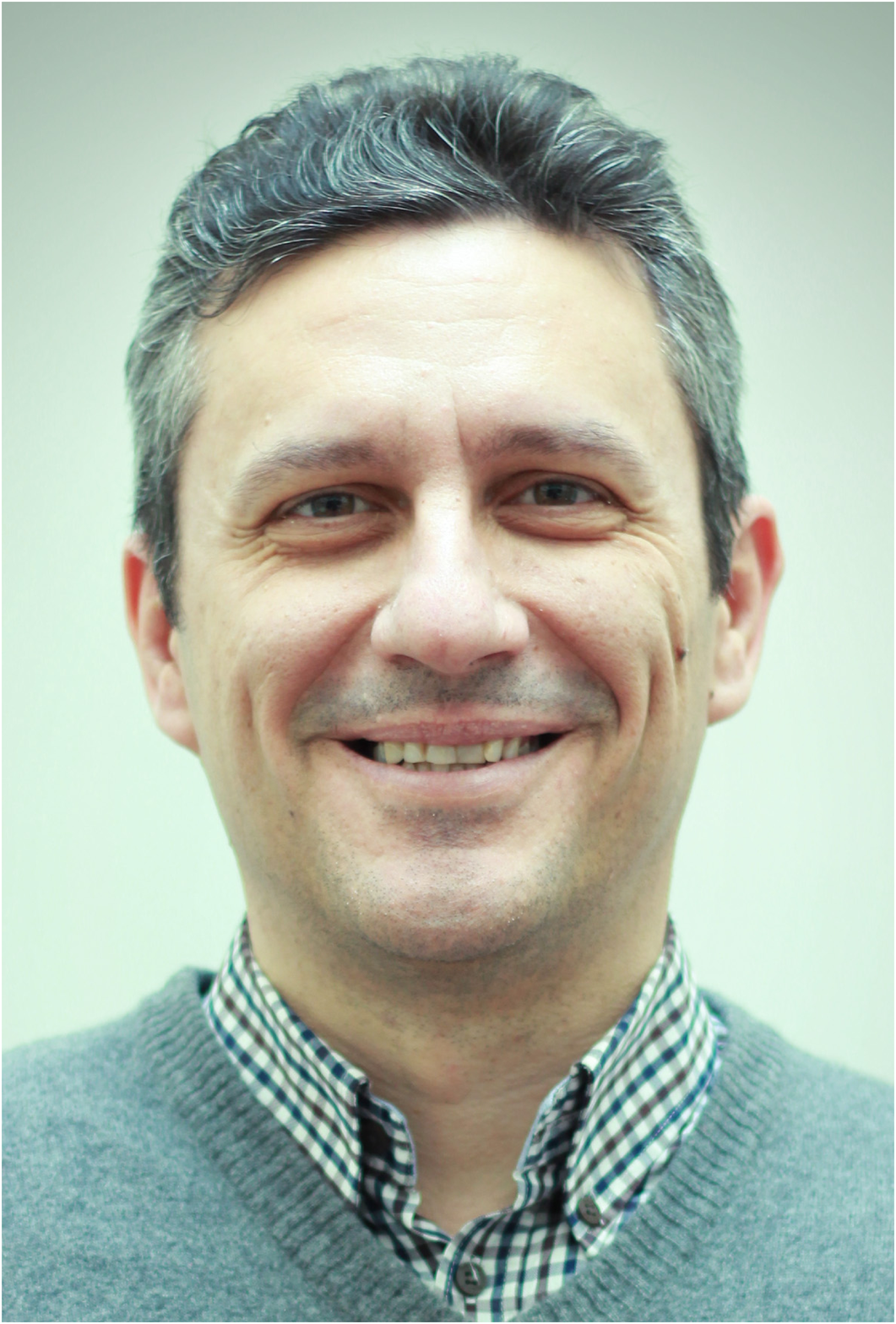}}]{Theodoros A. Tsiftsis} (S'02, M'04, SM'10) was born in Lamia, Greece, in 1970. He received the BSc degree in physics from the Aristotle University of Thessaloniki, Greece, in 1993, the MSc degree in digital systems engineering from the Heriot-Watt University, Edinburgh, U.K., in 1995, the MSc degree in decision sciences from the Athens University of Economics and Business, in 2000, and the PhD degree in electrical engineering from the University of Patras, Greece, in 2006. He is currently a Professor in the School of Intelligent Systems Science \& Engineering at Jinan University, Zhuhai, China, and also Honorary Professor at Shandong Jiaotong University, Jinan, China. His research interests fall into the broad areas of communication theory and wireless communications with emphasis on energy efficient communications, smart surfaces, ultra-reliable and low-latency communication, and physical layer security. 
	
	Dr. Tsiftsis has served as Senior or Associate Editor in the Editorial Boards of \textsc{IEEE Transactions on Vehicular Technology}, \textsc{IEEE Communications Letters}, \textsc{IET Communications}, and \textsc{IEICE Transactions on Communications}. He is currently an Area Editor for Wireless Communications II of the \textsc{IEEE Transactions on Communications} and an Associate Editor of the \textsc{IEEE Transactions on Mobile Computing}. Prof. Tsiftsis has been appointed to a 2-year term as an IEEE Vehicular Technology Society Distinguished Lecturer (IEEE VTS DL), Class 2018.  
\end{IEEEbiography}

\begin{IEEEbiography}
	[{\includegraphics[width=1in, height=1.25in, clip, keepaspectratio]{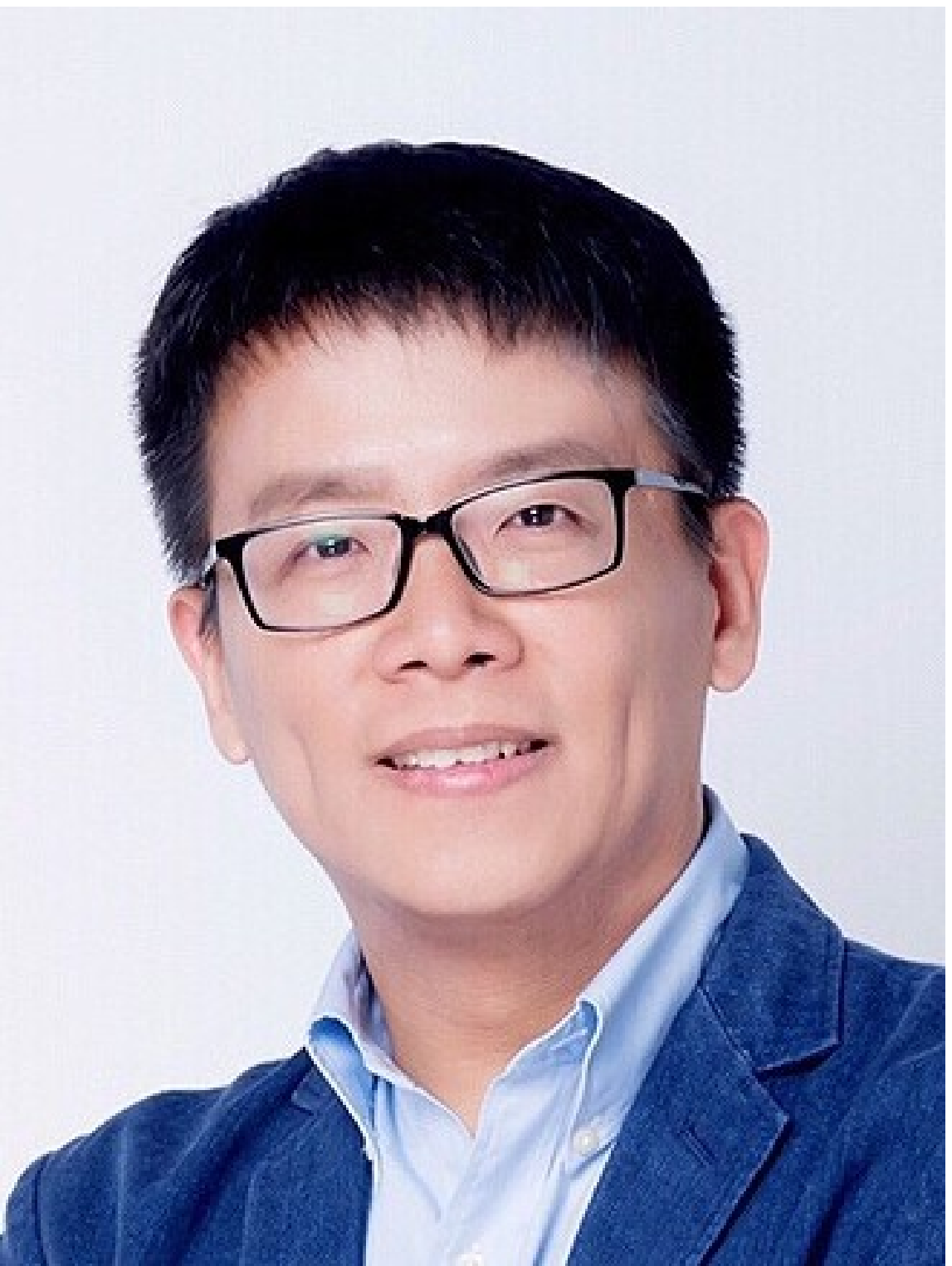}}]{Guanghua Yang} (S'02, M'07, SM'19) received his Ph.D. degree in electrical and electronic engineering from the University of Hong Kong, Hong Kong, in 2006. From 2006 to 2013, he served as post-doctoral fellow, research associate, and project manager in the University of Hong Kong. Since April 2017, he is an Associate Professor and Vice Dean with the School of Intelligent Systems and Engineering, Jinan University, Guangdong, China. His research interests are in the general areas of communications, networking and multimedia.
\end{IEEEbiography}

\begin{IEEEbiography}
	[{\includegraphics[width=1in, height=1.25in, clip, keepaspectratio]{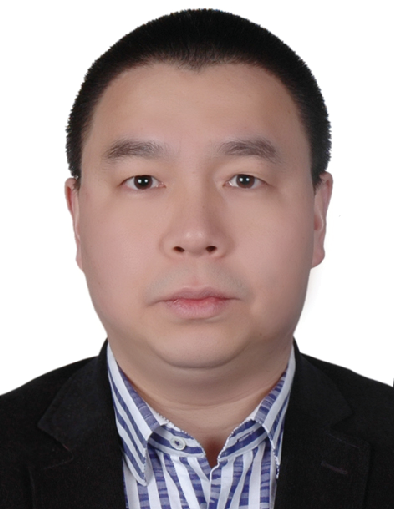}}]{Minghua Xia} (M'12) received the Ph.D. degree in Telecommunications and Information Systems from Sun Yat-sen University, Guangzhou, China, in 2007. 
	
	From 2007 to 2009, he was with the Electronics and Telecommunications Research Institute (ETRI) of South Korea, Beijing R\&D Center, Beijing, China, where he worked as a member and then as a senior member of engineering staff. From 2010 to 2014, he was in sequence with The University of Hong Kong, Hong Kong, China; King Abdullah University of Science and Technology, Jeddah, Saudi Arabia; and the Institut National de la Recherche Scientifique (INRS), University of Quebec, Montreal, Canada, as a Postdoctoral Fellow. Since 2015, he has been a Professor with Sun Yat-sen University. Since 2019, he has also been an Adjunct Professor with the Southern Marine Science and Engineering Guangdong Laboratory (Zhuhai). His research interests are in the general areas of wireless communications and signal processing. 
	
	Dr. Xia received the Professional Award at the IEEE TENCON, held in Macau, in 2015. He served as a TPC Symposium Chair of IEEE ICC'2019. He currently serves as a TPC Symposium Chair of IEEE ICC'2021, and an Associate Editor for the {\scshape IEEE Transactions on Cognitive Communications and Networking} and the {\it IET Smart Cities}. He was recognized as an Exemplary Reviewer by {\scshape IEEE Transactions on Communications} in 2014, {\scshape IEEE Communications Letters} in 2014, and {\scshape IEEE Wireless Communications Letters} in 2014 and 2015. 
\end{IEEEbiography}

\vfill

\end{document}